\documentclass[aps,prxquantum,reprint,nofootinbib,superscriptaddress]{revtex4-2}

\usepackage{amsmath,amssymb,bbm,dsfont,physics,upgreek}
\usepackage{hyperref}
\hypersetup{
	colorlinks,
	bookmarksopen = true,
	bookmarksnumbered = true,
	citecolor=blue,
	linkcolor=blue,
	urlcolor=blue
 	}
\usepackage{bigstrut}
\usepackage{booktabs}
\usepackage{etoolbox}
\usepackage{graphicx}
\usepackage{pgf,tikz}
\usepackage{pgfplots}
\usepackage{pgfplotstable}
\pgfplotsset{compat=1.7}
\usetikzlibrary{arrows,calc,math,fpu}
\pgfplotsset{compat=newest}
\usetikzlibrary{external}
\tikzset{
connection1/.style={densely dashed,black},
connection2/.style={densely dashed,brown}
}

\begin{document}
\newcommand{\ii}{\mathrm{i}}
\newcommand{\N}{\mathcal{N}}

\title{Sideband thermometry of ion crystals}

\author{Ivan Vybornyi}
\thanks{These three authors contributed equally.}
\affiliation{Institut für Theoretische Physik, Leibniz Universität Hannover, Appelstrasse 2, 30167 Hannover, Germany}

\author{Laura S. Dreissen}
\thanks{These three authors contributed equally.}
\affiliation{Physikalisch-Technische Bundesanstalt (PTB), Bundesallee 100, 38116 Braunschweig, Germany}
\affiliation{Department of Physics and Astronomy, LaserLab, Vrije Universiteit, De Boelelaan 1081, 1081 HV Amsterdam, The Netherlands}

\author{Dominik Kiesenhofer}
\thanks{These three authors contributed equally.}
\affiliation{Universität Innsbruck, Institut für Experimentalphysik, Technikerstraße 25, 6020 Innsbruck, Austria}
\affiliation{Institut für Quantenoptik und Quanteninformation, Österreichische Akademie der Wissenschaften, Technikerstraße 21a, 6020 Innsbruck, Austria}

\author{Helene Hainzer}
\affiliation{Universität Innsbruck, Institut für Experimentalphysik, Technikerstraße 25, 6020 Innsbruck, Austria}
\affiliation{Institut für Quantenoptik und Quanteninformation, Österreichische Akademie der Wissenschaften, Technikerstraße 21a, 6020 Innsbruck, Austria}

\author{Matthias Bock}
\affiliation{Universität Innsbruck, Institut für Experimentalphysik, Technikerstraße 25, 6020 Innsbruck, Austria}
\affiliation{Institut für Quantenoptik und Quanteninformation, Österreichische Akademie der Wissenschaften, Technikerstraße 21a, 6020 Innsbruck, Austria}

\author{Tuomas Ollikainen}
\affiliation{Universität Innsbruck, Institut für Experimentalphysik, Technikerstraße 25, 6020 Innsbruck, Austria}
\affiliation{Institut für Quantenoptik und Quanteninformation, Österreichische Akademie der Wissenschaften, Technikerstraße 21a, 6020 Innsbruck, Austria}

\author{Daniel Vadlejch}
\affiliation{Physikalisch-Technische Bundesanstalt (PTB), Bundesallee 100, 38116 Braunschweig, Germany}

\author{Christian F. Roos}
\affiliation{Universität Innsbruck, Institut für Experimentalphysik, Technikerstraße 25, 6020 Innsbruck, Austria}
\affiliation{Institut für Quantenoptik und Quanteninformation, Österreichische Akademie der Wissenschaften, Technikerstraße 21a, 6020 Innsbruck, Austria}

\author{Tanja E. Mehlstäubler}
\affiliation{Physikalisch-Technische Bundesanstalt (PTB), Bundesallee 100, 38116 Braunschweig, Germany}
\affiliation{Institut für Quantenoptik, Leibniz Universität Hannover, Welfengarten 1, 30167 Hannover, Germany}

\author{Klemens Hammerer}
\affiliation{Institut für Theoretische Physik, Leibniz Universität Hannover, Appelstrasse 2, 30167 Hannover, Germany}

\date\today

\begin{abstract}
    Coulomb crystals of cold trapped ions are a leading platform for the realisation of quantum processors and quantum simulations and, in quantum metrology, for the construction of optical atomic clocks and for fundamental tests of the Standard Model. For these applications, it is not only essential to cool the ion crystal in all its degrees of freedom down to the quantum ground state, but also to be able to determine its temperature with a high accuracy. However, when a large ground-state cooled crystal is interrogated for thermometry, complex many-body interactions take place, making it challenging to accurately estimate the temperature with established techniques. In this work we present a new thermometry method tailored for ion crystals. The method is applicable to all normal modes of motion and does not suffer from a computational bottleneck when applied to large ion crystals. We test the temperature estimate with two experiments, namely with a 1D linear chain of 4 ions and a 2D crystal of 19 ions and verify the results, where possible, using other methods. The results show that the new method is an accurate and efficient tool for thermometry of ion crystals.
\end{abstract}

\maketitle

\section{Introduction}
Trapped ions are one of the leading platforms for quantum computation~\cite{qc_rev,qc_qccd,qc_blpt}, simulation \cite{kielpinski_architecture_2002,Wright_2019,blatt_quantum_2012, monroe_programmable_2021}, sensing \cite{biercuk_ultrasensitive_2010,degen_quantum_2017}, and metrology~\cite{Keller2019,brewer_quantum-logic_2019,huntemann_single-ion_2016}. The exquisite degree of quantum control over electronic and motional degrees of freedom and their unrivalled coherence times make trapped ions excellent qubits~\cite{qc_rev,Wright_2019} and enable quantum gates with record fidelity~\cite{ballance_high-fidelity_2016,christensen_high_2020,srinivas_high_2021}. The unique combination of isolation and controllability also guarantees low systematic effects in precision measurements of the electronic structure of trapped ions, making them perfect systems for frequency and time references in next-generation optical clocks and for searches of physics beyond the Standard Model~\cite{safranova_search_2018}.

\begin{figure}[tbp]
    \centering
    \includegraphics{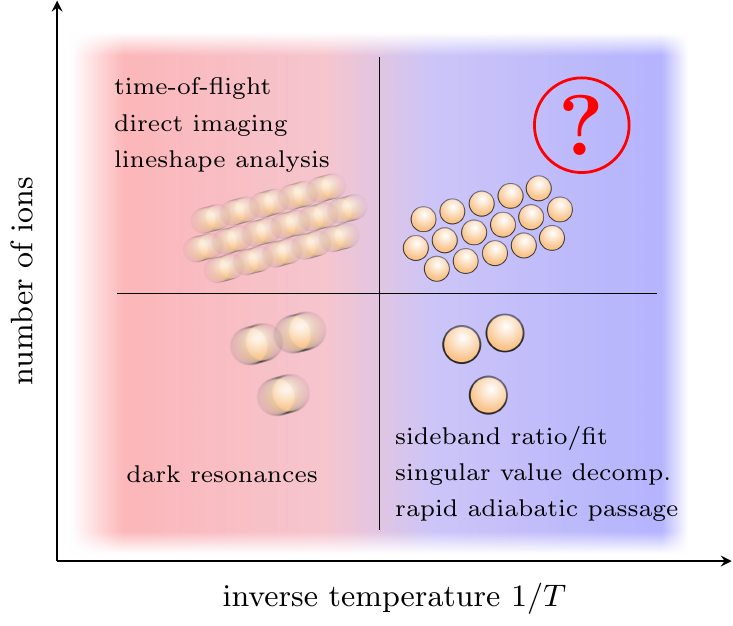}
    \caption{Schematic overview of thermometry methods for cold ions. Well-established approaches exist for ion crystals at large temperatures. Close to the ground state, the temperature can be measured of single ions or small crystals. Development of methods applicable to large and cold crystals is an open challenge and subject of the present work.}
    \label{fig:schema}
\end{figure}

Scaling up ion Coulomb crystals (ICCs) is desirable for all these applications, but poses increasing challenges in entropy management as the number of degrees of freedom of motion grows. This concerns first the efficient cooling of ICCs, since thermal excitations are a major limiting factor, e.g. in quantum gates mediated by normal modes of motion and, due to the relativistic Doppler effect, also in the systematics of ion clocks~\cite{kalincev_motional_2021,doppler_clocks}. Laser cooling of trapped ions has become a rich methodology~\cite{Eschner:03}, which in recent years has made it possible to cool even large ICCs consisting of dozens~\cite{Joshi_2020,PhysRevA.102.043110,PhysRevLett.125.053001} and even hundreds~\cite{PhysRevLett.122.053603} of ions near their ground states of motion. Beyond cooling, an equally significant challenge is to accurately measure the final state of motion of ICCs achieved by a particular cooling method and to determine its precise effective temperature. Accurate and efficient thermometry is important for evaluating spurious effects in quantum technology, such as quantum gate errors or systematic clock shifts, as well as for evaluating the efficiency of cooling schemes. 

There exist well-established methods for thermometry of trapped ions, yielding reliable results for ion Coulomb crystals (ICCs) in high-temperature thermal states and for single or few ions near the ground state, as summarized in Fig.~\ref{fig:schema}. Far from the motional ground state, time-of-flight, Doppler lineshape or image analyses achieve satisfactory accuracy~\cite{app10155264,Ro_nagel_2015,doi:10.1255/ejms.1408,PhysRevA.76.012719,Herrmann2009} and suit practically any ion numbers. Close to the motional ground state, the excellent control over the coupling of motional degrees of freedom to internal levels, combined with the extreme precision that can be achieved in measuring the latter, can be used to perform an indirect temperature measurement. For single trapped ions, techniques such as singular value decomposition and numerical fits are employed to analyze motional sideband transitions \cite{Meekhof1996,Rasmusson:2021}. The primary tool for single ion thermometry is the resolved sideband ratio technique \cite{Wineland:1987}, which exploits the pronounced asymmetry of motional sidebands near the ground state.

\begin{figure}[t]
    \centering
    \includegraphics{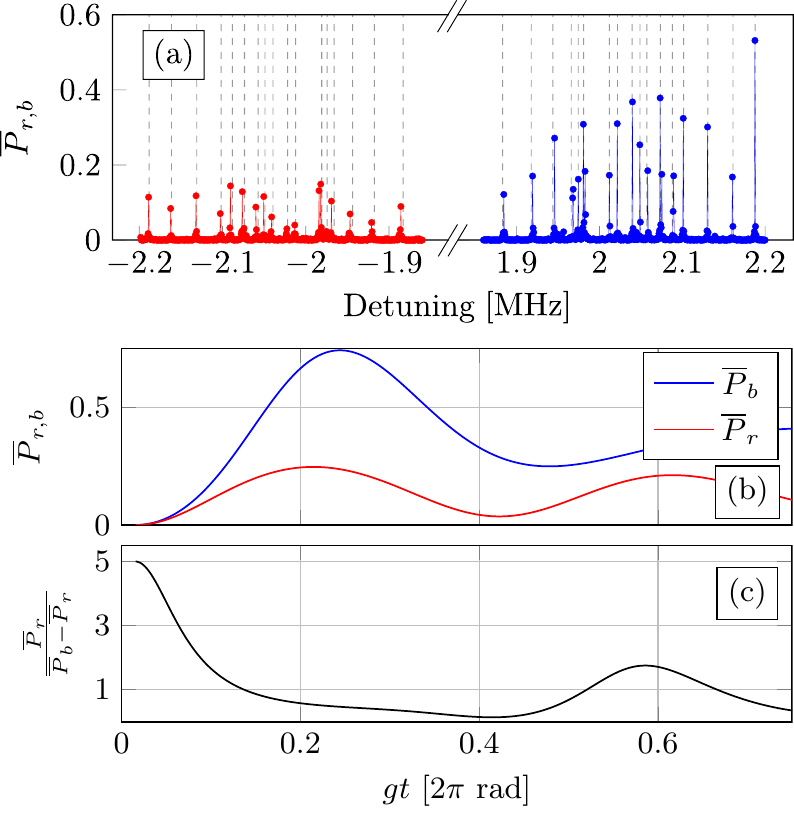}
    \caption[format=plain,justification=justified,singlelinecheck=false]{(a) Red and blue sideband spectrum of the out-of-plane modes of a 19-ion planar crystal after Doppler cooling (for further experimental details, see section~\ref{sec:19ion_thermometry}). $\overline{P}_{r,b}=\frac{1}{N}\sum_{i=1}^N P^{i}_{r,b} $ is the mean excitation probability averaged over the ion crystal under red and blue sideband drive, respectively. The absorption probability is much smaller on the red vibrational sidebands than on the blue sidebands. Yet, the motional modes are still far from being cooled close to their ground states with temperatures ranging at about 5-10 phonons on average. (b) Simulated dynamics of the mean electronic red/blue sideband excitation $\overline{P}_{r,b}$ of a 19-ion ICC (center-of-mass mode, assuming $\bar{n}=5$). (c) Taking $\overline{P}_r/(\overline{P}_b-\overline{P}_r)$ as an estimate of the mean phonon number, as suggested by the sideband ratio method (Eq.~\eqref{eq:sbratio}), yields completely erroneous results and a significant underestimate of the motion temperature.}
    \label{fig:19_spectrum_red_blue}
\end{figure}

However, these techniques cannot be directly applied to large ICCs near the ground state, posing an open challenge for achieving reliable thermometry in this regime, cf. Fig.~\ref{fig:schema}. The problem is that for globally addressed crystals, driving a first-order (red or blue) sideband transition induces strong and complex spin-spin correlations due to their joint coupling with the driven normal mode, much like in a trapped ion quantum gate. These correlations significantly affect the asymmetry between red and blue sideband transitions, as is illustrated in Fig.~\ref{fig:19_spectrum_red_blue} at the example of the sideband spectrum of a Doppler-cooled 19-ion crystal. The marked asymmetry in the mean excitation probabilities on the red and blue sidebands would naïvely suggest a mean phonon number of approximately $\bar{n}\lesssim 1$ when applying the sideband absorption ratio technique \cite{Wineland:1987} (cf. Eq.~\eqref{eq:sbratio}) to this scenario. However, this estimate significantly underestimates the actual mean phonon number, expected to be around $5$ to $10$ after Doppler cooling. This clearly shows that spin-spin correlations must be accounted for to accurately determine the temperature of motional states based on a measurement of internal state populations. For small ion crystals these correlations can be described exactly by solving the dynamics numerically or analytically \cite{home2006entanglement,doi:10.1080/09500340.2017.1376719,PhysRevLett.126.023604}. However for large ICCs consisting of tens or hundreds of ions, an exact consideration of ion-ion correlations is numerically intractable, as in any quantum many-body problem. Despite this complication, the resolved sideband technique (as well as other techniques based on spin-motional coupling) has been used in recent experiments~\cite{PhysRevLett.108.213003,PhysRevLett.127.020503,PhysRevLett.125.053001,PhysRevA.102.043110} where many-body correlations have been either partially or completely neglected.

The occurrence of many-particle dynamics in sideband thermometry can be trivially suppressed if only a single ion in a crystal can be interrogated. In this case, the known single-ion techniques are generally applicable, but become inefficient for larger ICCs due to poor statistics and long interrogation times required. When many or all ions of an ICC are interrogated, many-body dynamics can be easily accommodated in the exceptional case where the symmetric centre-of-mass (COM) mode is interrogated~\cite{IVANOV2019101,PhysRevLett.122.053603,PhysRevResearch.3.013244}. In this case, the dynamics preserves permutation symmetry and shows a growth of the effective dimension of the Hilbert space that is only polynomial, instead of exponential, in the number of ions. Accurate thermometry of the out-of-phase modes, i.e.~all modes except the COM mode, remains generally out of reach.

In this work, we present a new method for the thermometry of states of motion for arbitrarily large, globally addressed ICCs cooled close to the ground state. The method is based on measuring the ratio of the first-order excitation probabilities of the red and blue motional sidebands. Our technique takes into account the entanglement between the individual spins and remains accurate regardless of the number of ions. Using this global sideband thermometry method, we can extract the temperature of each motional mode that is assumed to be in a thermal motional state at the end of the cooling cycle. Our approach to the many-body aspect of the problem may also serve as a useful reference and possibly complement the other existing thermometry methods. We demonstrate our method on a linear four-ion ICC and verify the estimated result with a full numerical simulation. We also test the global sideband thermometry on a two-dimensional 19-ion ICC; in this case the comparison with the numerical simulation is only possible for the COM mode.

The article is organised as follows: We start with a theoretical description of the resolved sideband thermometry of a single ion in section \ref{subsec:theory_single_ion} and place it in the framework of parameter estimation. In section \ref{subsec:theory_ion_cryst} we present our new global sideband therometry method for ICCs and discuss its advantages and limitations. In section \ref{subs:loss} we describe an alternative bichromatic technique that could be used if single ion addressing is possible and compare it with our new method. In section \ref{sec:4ion_thermometry} we demonstrate the new global sideband method on four modes of motion of a linear 4-ion crystal and verify the results. In section \ref{sec:19ion_thermometry} we apply the new technique to a 2D 19-ion crystal and verify it. We summarise the results and give an outlook on open questions and possible further developments in section \ref{sec:disc}.

\section{Theory of sideband thermometry}
\label{sec:theory_sb_thermometry}
\subsection{Sideband Thermometry of a Single Ion}
\label{subsec:theory_single_ion}
\begin{figure}[tbp]
    \includegraphics{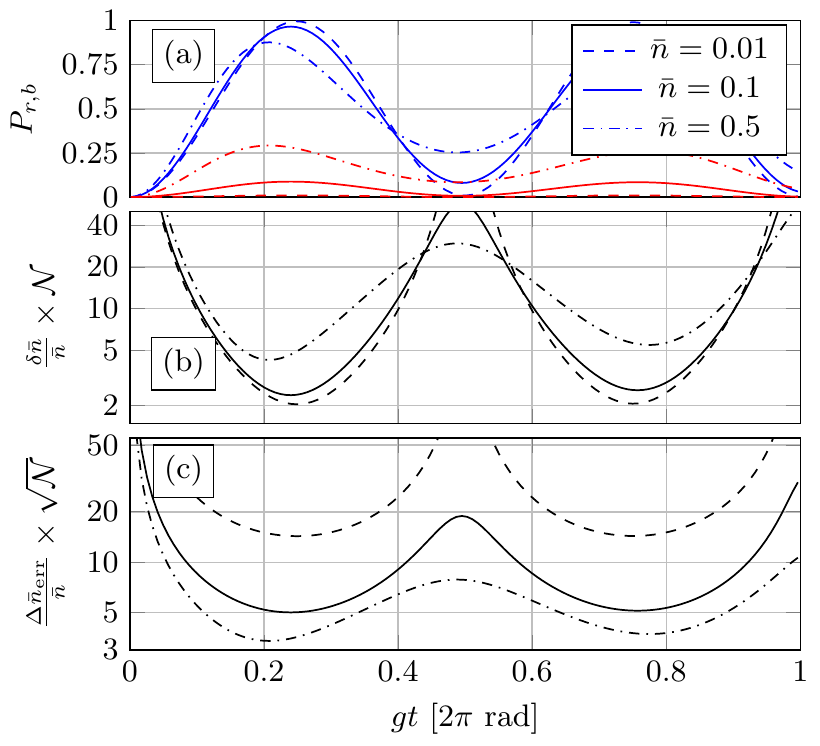}
    \caption[format=plain,justification=justified,singlelinecheck=false]{(a) A simulated Rabi flop on blue (blue curves) and red (red curves) sideband for a single ion for several values of the mean phonon occupation number. Bias (b) and the statistical uncertainty (c) of the sideband temperature estimator (\ref{eq:sbratio}), rescaled to the total number of measurements $\N$.}
    \label{fig:crb}
\end{figure}

We consider a trapped ion with two relevant internal states $\ket{\downarrow}$ and $\ket{\uparrow}$, harmonically bound in a quadrupole trap with quantized motion along the three main trap axes. We restrict ourselves to a single axis and assume the ion is prepared in a thermal state of motion $\rho(\bar{n})=\sum_{n=0}^\infty p_n(\bar{n})\ketbra{n}{n}$ with Fock state occupation probability
\begin{align}\label{eq:thermoccup}
p_n(\bar{n})=\frac{\bar{n}^n}{(\bar{n}+1)^{n+1}}.    
\end{align}
The mean occupation number is $\bar{n}=\Tr[\rho(\bar{n})a^\dagger a]$ where $a$ and $a^\dagger$ are bosonic creation and annihilation operators for the motional degree of freedom. The mean occupation number can be inferred by coupling the motional state to the internal states and measuring the excitation probability. This can be realized by initializing the ion in $\rho(\bar{n})\otimes\ketbra{\downarrow}{\downarrow}$ and illuminating it for a time $t$ by a laser driving resonant transitions on either the red or blue sideband. The dynamics in these cases is governed by the Hamiltonians
\begin{subequations}
    \begin{equation}
    H_r=g(\sigma_+a+\mathrm{h.c.})\,,
    \end{equation}
    \begin{equation}
    H_b=g(\sigma_+a^\dagger+\mathrm{h.c.})\,
    \end{equation}
\end{subequations}
respectively. Here, the effective Rabi frequency $g=\Omega \eta/2$ is obtained by rescaling the carrier Rabi frequency $\Omega$ with the Lamb-Dicke factor $\eta \ll 1$ and the spin operators are defined as $\sigma_+=\ketbra{\uparrow}{\downarrow}$. The probability to find the ion in the excited state $\ket{\uparrow}$ after an interrogation time $t$ is given by $P_{\alpha}(\bar{n},t)=\Tr[U_{\alpha}(t)\rho(\bar{n})\otimes\ketbra{\downarrow}{\downarrow} U^\dagger_{\alpha}(t)\ketbra{\uparrow}{\uparrow}]$, where $\alpha=r,b$ for red or blue sideband dynamics, respectively, and the time evolution operators are given by $U_{\alpha}(t)=\exp(-\mathrm{i}H_{\alpha} t)$. For convenience, the excitation probabilities are given explicitly in Eq.~(\ref{eq:flops_singleion}). Several examples of $P_{\alpha}(t,\bar{n})$ are plotted at specific values of $\bar{n}$ in Fig.~\ref{fig:crb}(a). 

A measurement of the excitation probability on both the red and blue sideband transitions as a function of interrogation time gives access to the mean occupation number via, e.g., a numerical fit of the data to $P_{\alpha}(\bar{n},t)$, \cite{PhysRevLett.126.023604}. Apart from this,
one can also infer the temperature from measurements at a single interrogation time by using the convenient formula
\begin{align}
    \frac{P_r(\bar{n},t)}{P_b(\bar{n},t)-P_r(\bar{n},t)}=\bar{n}\,.
    \label{eq:sbratio}
\end{align}
This identity is at the heart of the sideband ratio thermometry method  \cite{Wineland:1987,wineland_experimental_1998,leibfried_quantum_2003}. It can be easily verfied using the expressions for $P_\alpha(\bar{n},t)$ in Eq.~\eqref{eq:flops_singleion} and the fact that $\frac{p_{n+1}(\bar{n})}{p_n(\bar{n})}=\frac{\bar{n}}{\bar{n}+1}$ for thermal states. Eq.~(\ref{eq:sbratio}) is formally a correct analytical statement relating the measured excitation probabilities to the motional temperature. It is, however, important to note that this formula implicitly suggests that the values of $P_{\alpha}$ are the precise statistical averages and hence only holds true if the data comes from an infinitely large measurement sample.
In reality, the sample size is finite. The real measurement outcomes are the relative statistical frequencies $f_{\alpha}$, which will inevitably deviate from the true excitation probabilities. As the sample size $\mathcal{N}$ increases, this deviation goes down following a Gaussian law:
\begin{align}
f_{\alpha}\xrightarrow{\N\xrightarrow{}\infty}P_{\alpha}+\sqrt{\frac{2}{\N}}\sqrt{P_{\alpha}(1-P_{\alpha})}\xi_{\alpha},
\label{eq:sampling}
\end{align}
where $\xi_{\alpha}$ is a normally distributed random variable, $\xi_{\alpha} \sim N(0,1)$. 
Since $f_{\alpha}$ slightly differ from the real excitation probabilities, plugging these values into Eq.~(\ref{eq:sbratio}) results in an expression which is not exactly the motional temperature $\bar{n}$, but its statistical 'estimator':
\begin{align}\label{eq:estimator1}
\hat{\bar{n}}=\frac{f_r}{f_b-f_r}.
\end{align}
The values of the estimator (referred to as estimates for short) are distributed with a certain statistical uncertainty and might carry a bias, which has to be accounted for when applying Eq.~(\ref{eq:sbratio}) to experimental data. Using Eq.~(\ref{eq:sampling}) we calculate the statistical bias and the variance of the estimates:
\begin{subequations}
    \begin{align}
    \label{eq:bias_1ion}
        \delta \bar{n}&=\langle \hat{\bar{n}} \rangle -\bar{n} =\frac{1}{\N}\frac{2P_bP_r(2-P_b-P_r)}{(P_b-P_r)^3},\\
    \label{eq:dn_1ion}
        \Delta\bar{n}_{\textrm{err}}^2&=\langle (\hat{\bar{n}}-\bar{n})^2\rangle = \frac{1}{\N}\frac{2P_bP_r(P_b+P_r-2P_bP_r)}{(P_b-P_r)^4}\,.
    \end{align}
\end{subequations}

Both the variance and the bias converge to zero with the number of measurements. In Fig.~\ref{fig:crb}(b, c) the intrinsic bias and the relative statistical uncertainty of the estimator in Eq.~(\ref{eq:sbratio})  are plotted as a function of the interrogation time and rescaled to $\N$. As one would expect, both bias and the statistical uncertainty are large at small interrogation times, where the obtained statistics is poor due to the weak distinguishability of the sidebands. 
The bias and uncertainty also strongly scale with mean occupation number $\bar{n}$. For example, achieving an uncertainty of $3\%$ at $\bar{n}=0.5$ requires $\mathcal{N}=10^4$ measurements at an optimal interrogation time.
An optimal interrogation time is found near the time of the maximal sideband excitation. This is the operating point, where the estimation using Eq.~(\ref{eq:sbratio}) converges the fastest to the true temperature value and has minimal bias. It is also interesting to compare the statistical uncertainty of the estimator with the fundamental bounds given by the (quantum) Cramér-Rao bounds. Since these observations are not central to the thermometry problem we consider below, we defer them to Appendix~\ref{app:FIzeros}.

\subsection{Sideband Thermometry of an Ion Crystal}
\label{subsec:theory_ion_cryst}
In an ICC, the motion of ions is collective and occurs in normal modes~\cite{wineland_experimental_1998}. In the following, we operate under the commonly used premise that after laser cooling, each mode is in a thermal state characterised by a certain mean phonon number $\bar{n}$. This is usually a good approximation, although there may well be deviations, especially for short cooling times~\cite{Chen:2017,Rasmusson:2021,Kulosa2023}. Performing ion crystal thermometry thus amounts to estimating the mean phonon number of the collective modes of motion by performing spin measurements on the crystal. 

To date, many of the used temperature estimation techniques operating in the nearly ground-state cooled regime ($\bar{n}\lesssim 1$) rely on knowing the exact system dynamics governed by the first-order red and blue sideband Hamiltonians, arising in the Lamb-Dicke regime of the atom-light interaction for a globally-addressed crystal. These types of coupling entangle the crystal's particular mode of motion with the spins and are given by
    \begin{subequations}\label{eq:Hams}
    \begin{align}
    H_r&=g(J_+a+J_-a^{\dagger})\,,\label{eq:hamred}\\
    H_b&=g(J_+a^{\dagger}+J_-a)\,,\label{eq:hamblue}
    \end{align}
    \end{subequations}
with (pseudo-)collective spin operators $J_{\pm}=\sum_{i=1}^N\eta_i\sigma^i_{\pm}$, where $N$ is the number of ions in the crystal. Here, $g$ denotes an average Rabi frequency of sideband transitions. The average is chosen such as to have $\sum_i\eta^2_i=1$, where the factors $\eta_i$ account for all inhomogeneities in the couplings of the ions to the mode under consideration. Thus, $\eta_i$ comprises the net effect of Lamb-Dicke factors, varying Rabi-frequency etc., which we assume to be calibrated for a given setup.

To obtain the desired exact solution, one needs to time-propagate the Schrödinger equation with Hamiltonians (\ref{eq:Hams}). The complexity of this calculation grows exponentially with the number of ions $N$, making it impractical to extract the exact solution for large ion crystals in reasonable time. The only exceptional case is the symmetric center-of-mass mode, where all the individual coupling constants are equal, $\eta_i\equiv 1/\sqrt{N}$. This allows one to solve the problem in a symmetric Hilbert subspace, thus lifting the exponential scaling, cf. Appendix~\ref{app:rabi}.

In order to resolve the large-$N$ bottleneck for out-of-phase modes, a new temperature estimation method is needed, which does not require solving the Schrödinger equation for the dynamics of the coupled $N$-ion system. We intend to stay as close as possible to the concept of thermometry of a single ion, which we have described in Sec.~\ref{subsec:theory_single_ion}. To extend this approach, we first need to define the excitation probabilities for an ion crystal. A naive way to do this would be to use the mean electronic excitation of the ions, $\overline{P}_{\alpha}=\frac{1}{N}\sum_{i=1}^N P^{i}_{\alpha}$, where $P^{i}_{\alpha}=\Tr{\rho_{\alpha}\ket{\uparrow}_i\!\bra{\uparrow}}$. However, adopting this definition of the excitation probability for sideband thermometry leads to an incorrect result, and the temperature is greatly underestimated, as already shown in Fig.~\ref{fig:19_spectrum_red_blue}.

We will show that it is much more efficient to define the global crystal excitation probability as the complement of the probability to remain in the ground state of all ions, i.e.

\begin{align}
P_{\alpha}(\bar{n},t)=1-\Tr[\dyad{\mathbf{0}}{\mathbf{0}}\rho_{\alpha}(\bar{n},t)].\label{eq:pdef}
\end{align}
Here, $\ket{\mathbf{0}}=\ket{\downarrow\dots\downarrow}$ is the collective spin ground state, and $\rho_{\alpha}(\bar{n},t)=U_\alpha( t)\rho(\bar{n})\otimes\ketbra{\mathbf{0}}{\mathbf{0}}U^\dagger_\alpha(t)$ the time-propagated full density matrix of the system for red/blue sideband excitation. 

Defining the excitation probability in this way has two benefits: firstly, the measurement implied by Eq.~\eqref{eq:pdef} can be easily performed and does not require single ion resolution. Secondly, for sufficiently small $\bar{n}$ and $t$, the functional dependence between the sideband ratio and $\bar{n}$ can be computed efficiently, taking into account the entanglement between the ions in the crystal. The generalization of Eq.~\eqref{eq:sbratio} for ICCs is
\begin{subequations}\label{eq:mainresult}
    \begin{equation}
    \frac{P_r(\bar{n},t)}{P_b(\bar{n},t)-P_r(\bar{n},t)}=\mathcal{R}_t(\bar{n}),
    \label{eq:relation}
    \end{equation}
where
    \begin{equation}
    \mathcal{R}_t(\bar{n})=\bar{n} + (gt)^2\mathcal{P}_2(\bar{n})-(gt)^4\mathcal{P}_3(\bar{n}) + (gt)^6\mathcal{P}_4(\bar{n})+o(t^8)\,.
    \label{eq:maineq} 
    \end{equation}
\end{subequations}
The $\mathcal{P}_k(\bar{n})$ are certain known polynomials in $\bar{n}$ of order $k$ with coefficients depending solely on the mode coupling coefficients $\eta_i$. Their explicit form and details of their derivation can be found in Appendix~\ref{app:derivation}. Further below we will comment on the idea behind this calculation.

\begin{figure}[tbp]
    \centering
    \includegraphics{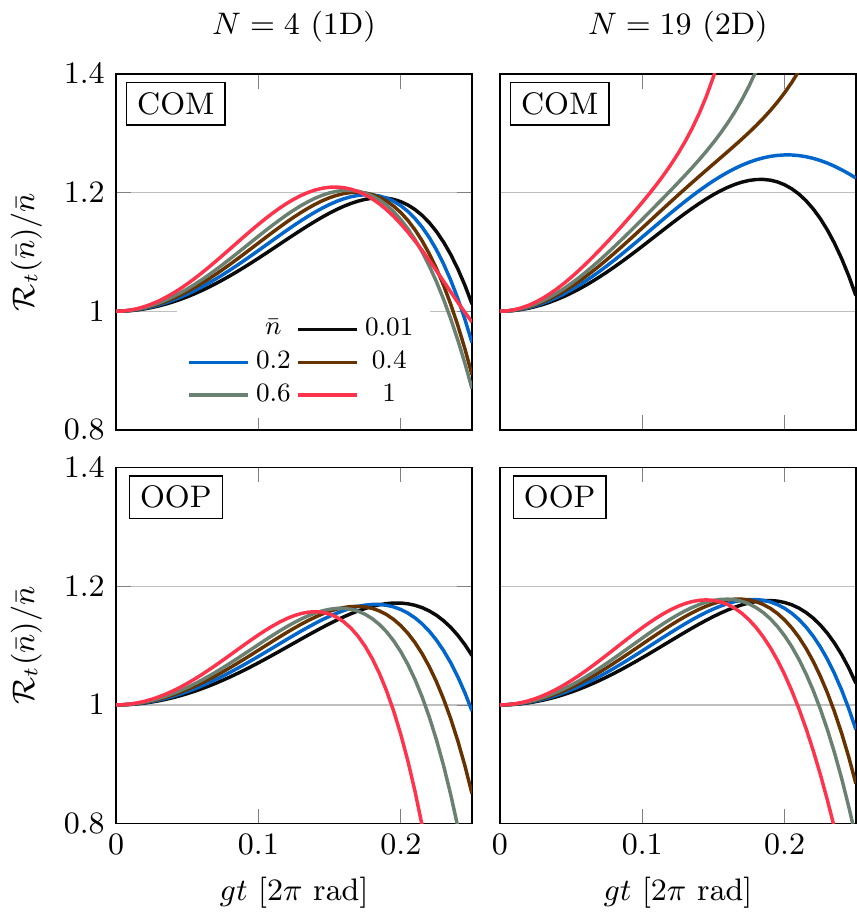}
    \caption{Example plots of $\mathcal{R}_t(\bar{n})/\bar{n}$ for several values of temperature. Considered are the COM-mode and one representative out-of-phase mode of a 1D 4-ion crystal and of a 2D 19-ion crystal, which are interrogated in Sections \ref{sec:4ion_thermometry} and \ref{sec:19ion_thermometry}.}
    \label{fig:roft}
\end{figure}

In Fig.~\ref{fig:roft} we plot $\mathcal{R}_t(\bar{n})/\bar{n}$ as a function of time for temperature values in the regime of interest. The curve's deviation from the values of 1 thus shows the relative temperature estimation error when naïvely applying the single-ion formula~\eqref{eq:sbratio} to an ICC. As case studies, we plot the curves for a 1D 4-ion crystal as considered in Section~\ref{sec:4ion_thermometry} and for a 2D 19-ion crystal considered in Section~\ref{sec:19ion_thermometry}. The chosen modes are the COM-mode and one representative out-of-phase mode. As is evident from Fig.~\ref{fig:roft},  applying formula~\eqref{eq:sbratio} to an ICC will result in a relative error of roughly 20\%, depending on the chosen mode and the crystal interrogated. Comparing this result with Fig.~\ref{fig:19_spectrum_red_blue}(b,c) shows that defining the global excitation probability as per (\ref{eq:pdef}) significantly improves the temperature estimation using the single-ion formula~\eqref{eq:sbratio} compared to using the mean electronic excitation. The new estimator in Eq.~\eqref{eq:mainresult} accounts for the residual discrepancy and thus mitigates the systematic error.

In the spirit of the thermometry of a single ion discussed in Sec.~\ref{subsec:theory_single_ion}, the temperature estimator appropriate for the normal mode of an ICC can be constructed by replacing the excitation probabilities on the left hand side of Eq.~\eqref{eq:mainresult} by measured relative statistical frequencies, and solving the resulting equation for $\bar{n}$. Thus, with the (properly chosen) root of the quartic polynomial $\mathcal{R}_t(\bar{n})-f_r/(f_b-f_r)=0$, the temperature estimator is
\begin{align}\label{eq:new_est}
    \hat{\bar{n}}=\mathcal{R}_t^{-1}\qty(\frac{f_r}{f_b-f_r}).
\end{align}
Eq.~\eqref{eq:new_est} is the sought for generalization of Eq.~\eqref{eq:estimator1} to an ICC. The systematic bias and the estimation error for finite sample size can be computed as in the case of a single ion, and are given in Eqs.~(\ref{eq:crystal_bias}) and (\ref{eq:crystal_var}) of the Appendix~\ref{app:derivation}. Only sideband excitation probabilities and $\mathcal{R}_t(\bar{n})$ with its derivatives enter the formula Eq.~(\ref{eq:crystal_bias}). Hence, no new data needs to be gathered to perform the bias correction.

Before discussing the properties and limitations of this estimator, we sketch how Eqs.~\eqref{eq:mainresult} are derived. Firstly, we exploit that both Hamiltonians in Eqs.~\eqref{eq:Hams} have a conserved quantity, namely $\comm{H_r}{a^\dagger a+J_0}=\comm{H_b}{a^\dagger a-J_0}=0$, where $J_0=\sum_{i=1}^N \sigma^i_+\sigma^i_-$ measures the  number of spin excitations. In both cases, $\alpha=r,b$, this entails for the probability in Eq.~\eqref{eq:pdef} of remaining in the spin ground state that
\begin{align}\label{eq:matelem}
    \Tr[\dyad{\mathbf{0}}{\mathbf{0}}\rho_{\alpha}(\bar{n},t)]=\sum_{n=0}^\infty p_n(\bar{n})\abs{\bra{\mathbf{0},n}U_\alpha(t)\ket{\mathbf{0},n}}^2.
\end{align}
where $p_n(\bar{n})$ is the thermal occupation probability according to Eq.~\eqref{eq:thermoccup}. Thus, only diagonal components of the time evolution operator enter the excitation probability $P_\alpha(\bar{n},t)$.

Secondly, we use that it is sufficient to describe the time dependence of the excitation probabilities up to their first fringe, as is evident from the discussion in Sec.~\ref{sec:theory_sb_thermometry}. This observation holds true for the ion crystal case as well. 
To exploit this, the diagonal matrix elements of the evolution operator are expanded in a power series in time. The series is then truncated at eighth order, since it is found to be enough to cover the significant fraction of the first sideband excitation fringe:
\begin{align}
    \bra{\mathbf{0},n}U_\alpha(t)\ket{\mathbf{0},n}\simeq \sum_{k=0}^4 \frac{(-\ii t)^{2k}}{(2k)!}\bra{\mathbf{0},n}H_{\alpha}^{2k}\ket{\mathbf{0},n}+o(t^{10}).
    \label{eq:series}
\end{align}
Note that only even powers of the sideband Hamiltonians make a non-vanishing contribution. The relevant matrix elements $\bra{\mathbf{0},n}H_{\alpha}^{2k}\ket{\mathbf{0},n}$ are polynomials in $n$ of degree $k$, and comprise the many-body dynamics generated by the sideband drive. The matrix elements can be evaluated analytically, and the lowest two orders are given in Appendix~\ref{app:derivation}. The expressions for the cases $k=3,4$ are rather lengthy, and are evaluated by means of computer algebra, which is available at~\cite{REPO}. With the approximation in Eq.~\eqref{eq:series}, the matrix elements of the evolution operator become polynomials of
order four in $n$ and eight in $gt$. The average with respect to $n$ in Eq.~\eqref{eq:matelem} can then be taken exactly, and yields the excitation probabilities in Eq.~\eqref{eq:pdef} as polynomials in $\bar{n}$ and $gt$, still of order four and eight, respectively. Since $P_{\alpha}(\bar{n},0)=0$, the final result for the ratio $\mathcal{R}_t(\bar{n})$ in Eq.~\eqref{eq:relation} is correct within sixth order in $gt$.

\begin{figure}[tbp]
    \includegraphics{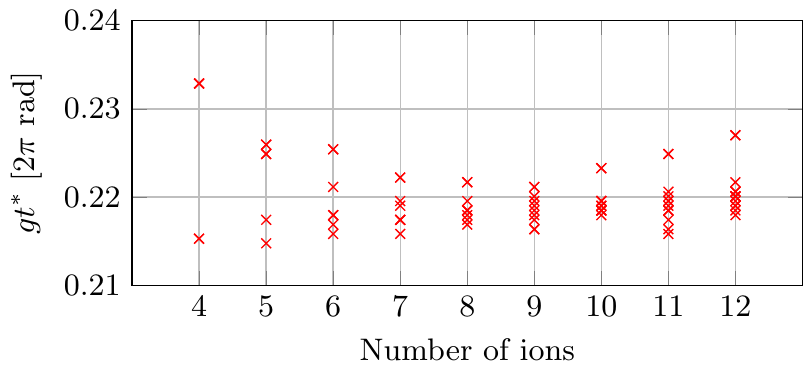}
    \caption{Cutoff time $gt^*$ for $N$ motional modes of a crystal containing $N=4$ to $12$ ions (for a linear chain with generic trap parameters). Each point corresponds to a particular motional mode. Due to the symmetries in the mode vectors, certain modes have coinciding cutoff time values. The upper points on the plot correspond to the COM mode. The data refers to $\bar{n}=0.1$.}
    \label{fig:cutoffs}
\end{figure}

Since the new estimator (\ref{eq:new_est}) relies on a time-truncation of the dynamics, cf.~Eq.~(\ref{eq:series}), the temperature estimation will be reliable only up to a certain time, which should cover a significant fraction of the first fringe in the excitation probabilities. In order to investigate this more quantitatively, we define a `cutoff time' $t^*$ at which the estimated mean phonon number deviates from the true value for more than a predefined error threshold $\epsilon$, which we chose to be $\epsilon=5\cdot10^{-3}$. For small ion crystals, $t^*$ can be calculated numerically and its dependence on the size of the ICC can be investigated. In Fig.~\ref{fig:cutoffs} the cutoff time is shown for all motional modes for a case study of ion crystals containing $N=4$ to $12$ ions. The results show that there is no significant dependence of $t^*$ on the motional mode index or on the number of ions, and that the cutoff time is sufficient to measure an excitation signal with good signal-to-noise ratio on both motional sidebands. We also observed no tendency for the cutoff time to significantly decrease when increasing the temperature within the regime $\bar{n}\lesssim 1$. Since no assumptions on the crystal size were made along the derivation, one may infer that the proposed global sideband temperature estimator of Eq.~(\ref{eq:new_est}) remains valid for large ion crystals. 

In summary, the estimator in Eq.~\eqref{eq:new_est} provides a suitable extension of the well-established sideband thermometry to large, cold ICCs. It allows collective addressing and readout of the ions,  providing fast dynamics and a strong signal, and adequately reflects the many-particle correlations involved. In the following, we will indicate yet another approach to ICC thermometry which also allows for collective addressing, but exploits single ion readout in order to avoid the complications connected to many-body dynamics. However, global sideband thermometry as discussed in Sec.~\ref{subsec:theory_ion_cryst} gives better statistics at low temperatures, as we will show.

\subsection{Thermometry from collective  bichromatic drive and single ion readout}\label{subs:loss}

When the red and blue sidebands are driven \emph{simultaneously}, the  dynamics of the ICC follows the Hamiltonian $H=H_r+H_b=\sum_i H^i$ with commuting single particle Hamiltonians $H^i=g\eta_i (a+a^\dagger)\sigma_x^i$. This can be exploited to avoid the difficulty of dealing with complex many-body interactions, when the readout can be done via a particular \emph{single} ion of the crystal. For a crystal initially prepared in the state $\rho_0=\ketbra{\mathbf{0}}{\mathbf{0}}\otimes\rho(\bar{n})$, the probability to find atom $i$ in the excited state after a time $t$ of bichromatic driving is
\begin{align}
    P^i(\bar{n},t)&=
    \Tr[e^{-\ii H^i t}\ketbra{\downarrow}{\downarrow}\otimes\rho(\bar{n}) e^{\ii H^i t}\ketbra{\uparrow}{\uparrow}]\nonumber\\
    &=\frac{1}{2}\qty(1-e^{-2(gt\eta_{k})^2(2\bar{n}+1)}).
    \label{eq:single_est_new}
\end{align}
This shows an exponential loss of contrast at a rate determined by the sought-after mean phonon number $\bar{n}$, independent of the exact dynamics of the other ions in the crystal. It can therefore form the basis for an estimator of the motion temperature without having to consider particle correlations. However, the necessary interrogation time will depend strongly on the chosen ion via the mode coefficient $\eta_i$, and might get large as the crystal size is increased. Moreover, interrogation of a single ion will require larger statistics.

\begin{figure}[tbp]
    \includegraphics{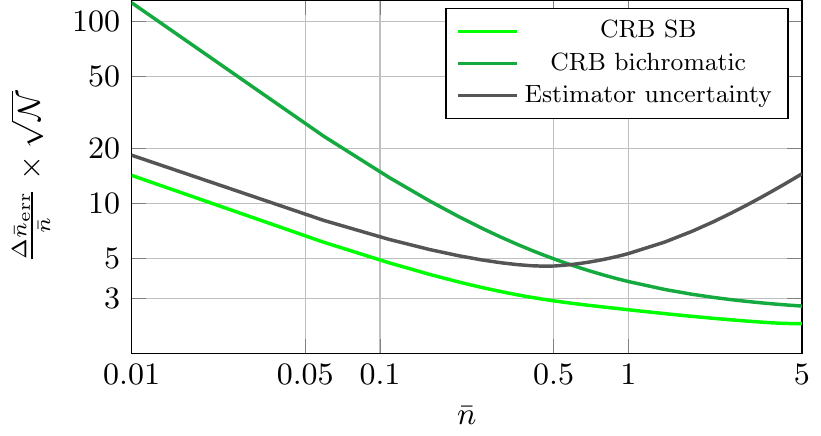}
    \caption[format=plain,justification=justified,singlelinecheck=false]{
    Cramér-Rao bound for temperature estimation in case of the sideband measurement on the whole crystal (bright green) and in case of the measurement based on bichromatic drive and interrogation of a single ion (dark green). For comparison, the statistical uncertainty of the global sideband temperature estimator (\ref{eq:new_est}) is plotted (black curve). Here $N=10$, the center-of-mass mode is considered. The curves for other modes overlap almost completely with the COM-curves.}
    \label{fig:compared}
\end{figure}

A quantitative comparison between thermometry based on bichromatic drive and the approach to global sideband thermometry can be obtained by considering their statistical uncertainties and their Cramér-Rao (CR) lower bounds. The latter follow directly from the probabilities in Eqs.~\eqref{eq:pdef} and \eqref{eq:single_est_new}, respectively. In each case, the interrogation time can be optimized to achieve minimal estimation error at a given motional mode temperature $\bar{n}$. For the bichromatic drive, the CR bound is independent of the number of ions and the specific mode. For the global sideband thermometry we use as an example the COM mode of a 10-ion crystal, and compare the CR bound also to the error for the specific estimator in Eq.~\eqref{eq:new_est}. The results are shown in Fig.~\ref{fig:compared} and demonstrate that the two methods show advantages in complementary regimes: for low temperatures, sideband thermometry will yield better statistics, while for higher temperatures corresponding to $\bar{n}\gtrsim 1$ the bichromatic approach is more efficient. The sideband ratio estimator does not saturate its CRB, yet it gets close to the CRB curve in the limit of small $\bar{n}$, while diverging from it for larger $\bar{n}$. This is mostly due to the fact that outside of the regime $\bar{n}\lesssim 1$ the cutoff time starts to decrease significantly with growing temperature, shifting the statistical uncertainty minima to higher values. As the cutoff time does not cover the optimal interrogation time required for the CR bound, there is space for getting closer to the bound with a higher cutoff time, which could be achieved by increasing the truncation order in~ \eqref{eq:series}. 

\section{Thermometry of a linear 4-ion crystal}
\label{sec:4ion_thermometry}
To test the new global sideband themometry method described in Sec.~\ref{subsec:theory_ion_cryst} we measure the motional temperature of a linear ICC of four $^{172}$Yb$^+$ ions. The size of the crystal allows us to evaluate the sideband dynamics exactly and thus benchmark the new method by comparing it to a direct numerical fit.

\subsection{State preparation and cooling}
The crystal is stored in a segmented linear radio-frequency (rf) Paul trap \cite{pyka_high-precision_2014,keller_probing_2019}. The radial confinement, i.e.~in the $xy$ plane, is set by an rf electric field driven at $\Omega_{\mathrm{rf}}=2 \pi \times 24.4$\,MHz, which is supplied to the trap electrodes by a resonant circuit. The axial confinement, i.e.~along $z$ axis, is set by a combination of dc voltages supplied to the trapping segment and both neighbouring segments. The corresponding secular frequencies are $\omega_{x,y,z}=2\pi\times (586,666,111)$\,kHz. The 12 motional modes along the $x$, $y$ and $z$ directions are cooled to the Doppler temperature of about $0.5$\,mK on the dipole allowed $^2\mathrm{S}_{1/2}\rightarrow$ $^2\mathrm{P}_{1/2}$ transition near 370\,nm, assisted by a repumper laser near 935\,nm. The ions are detected individually by collecting the fluorescence from the decay of the short-lived $^2\mathrm{P}_{1/2}$ state. With a high numerical-aperture lens of $\mathrm{N/A}=0.27$ the light from individual ions is imaged onto an electron multiplying charge-coupled device (EMCCD). For more details on the experimental apparatus, see \cite{pyka_high-precision_2013,pyka_high-precision_2014,kalincev_motional_2021}.

The four modes along the strong radial axis of the trap ($y$) are further cooled to near the ground state using quench-assisted resolved-sideband cooling on the $^2\mathrm{S}_{1/2}\rightarrow$ $^2\mathrm{D}_{5/2}$ electric quadrupole transition near 411\,nm and the dipole allowed $^2\mathrm{D}_{5/2}\rightarrow$ $^2\mathrm{P}_{3/2}$ near 1650\,nm. The 411\,nm beam is derived from a 822\,nm laser that is locked to a cavity with a fractional instability of $5\times10^{-16}$ at 10\,s of averaging time. It propagates parallel (within $3^{\circ}$) to the strong radial axis of the trap and efficiently addresses only the corresponding radial $y$-modes. It is focused down to a waist of 50\,$\mu$m at the position of the ions and is aligned to the center of the crystal by measuring the carrier Rabi frequency of all four ions individually. After optimization, the Rabi frequencies are measured to be $\Omega_{\mathrm{Rabi}}[\mathrm{ion} 1,\mathrm{ion} 2,\mathrm{ion} 3,\mathrm{ion} 4] = 2\pi\times[10.66(6),10.61(6),10.58(6),9.88(3)]$\,kHz, which varies by at most $10\%$ over the crystal. The exact mode frequencies of the four radial $y$-modes are measured to be $\omega_{y}[\mathrm{mode 1}, \mathrm{mode 2}, \mathrm{mode 3}, \mathrm{mode 4}]= 2\pi\times[666.0(1),656.9(1),643.1(1),623.6(1)]$\,kHz with resolved sideband spectroscopy on the 411\,nm transition. These mode frequencies are used to calculate the Lamb-Dicke factors for the motional modes, where the COM mode is at the highest mode frequency. In order to cool all modes simultaneously, the frequency of the 411\,nm laser is set to be $640$\,kHz red-detuned from the carrier transition, such that it is roughly at the center of the four measured mode frequencies. The 1650\,nm laser propagates along the axis of the trap and the power is tuned to reach an effective linewidth of the three level system of $67(2)\,\mathrm{kHz}$, see also \cite{Kulosa2023}. 

\subsection{Sideband thermometry measurement}
\begin{figure}[tbp]
    \centering
      \includegraphics{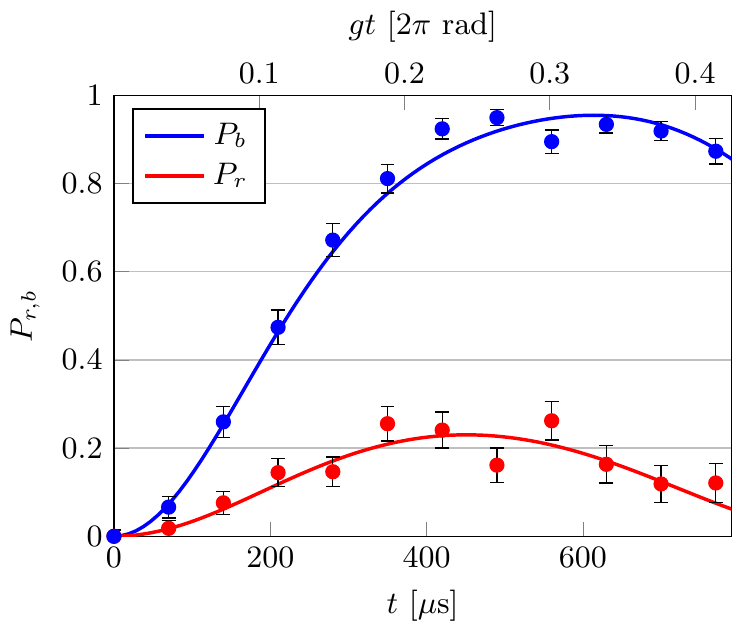}
     \caption{Red (red points) and blue (blue points) global sideband flops of the third motional mode in a 4-ion crystal (see the text for details). The lines show the exact numerical solution generated with the least-square fitted temperature value $\bar{n}$.}
    \label{fig:sbf_mode3}
\end{figure}

\begin{figure}[tbp]
    \centering
    \includegraphics{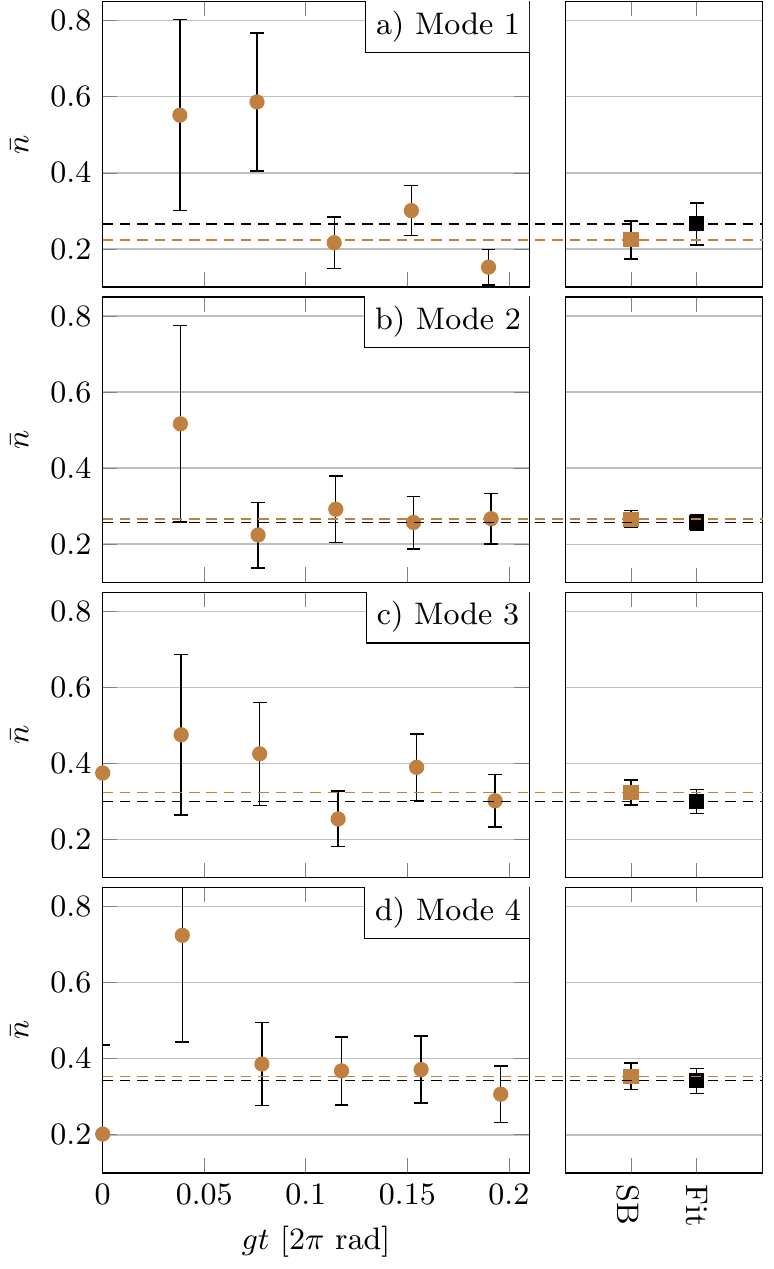}
    \caption{Temperature estimation for motional modes 1-4 shown in subpanels (a)-(d), respectively, of a 4-ion crystal. The individual temperature estimations at interrogation times up to the cutoff time of $\sim 400$\,$\mu$s (left panel) are averaged together to produce the final value of $\bar{n}=\{0.22\pm 0.05, 0.27\pm 0.02, 0.32 \pm 0.03, 0.35\pm 0.04\}$ shown in the right panel using the presented global sideband method. The estimations for the extremely small interrogation time of $10 \mu$s (the seen very left point for modes 3 and 4) and their uncertainties fall outside of the plot range. The results are compared to the values obtained from a least-square numerical fit from the `Rabi flops' of the red and blue sidebands. Agreement is found between the two methods within 1$\sigma$.}
    \label{fig:nbars}
\end{figure}

After ground-state cooling, a thermometry measurement is performed on each motional mode along the $y$ direction. The corresponding red and blue sidebands on the $^2\mathrm{S}_{1/2}\rightarrow$ $^2\mathrm{D}_{5/2}$ transition are addressed to measure the excitation probabilities $P_{r}$ and $P_{b}$, as defined in Sec.\,\ref{subsec:theory_ion_cryst}. For simplicity, the electronic states of individual ions are denoted as $\ket{^2\mathrm{S}_{1/2}}_{i}=\ket{\downarrow}_{i}$ and $\ket{^2\mathrm{D}_{5/2}}_{i}=\ket{\uparrow}_{i}$. The internal state of each individual ion is measured spatially resolved after a sideband pulse using the electron shelving technique, i.e.~fluorescence was only detected on the $^2\mathrm{S}_{1/2}\rightarrow ^2\mathrm{P}_{1/2}$ transition if the ion was in the $\ket{\downarrow}$ state. If $\ket{\downarrow}_{1}=\ket{\downarrow}_{2}=\ket{\downarrow}_{3}=\ket{\downarrow}_{4}$, $P_{\alpha}=0$, otherwise $P_{\alpha}=1$, where $\alpha =r,b$. For a specific interrogation time, 
the excitation probabilities $P_{r}$ and $P_{b}$ are obtained by averaging over $\N/2=200$ 
measurements and the interrogation time is scanned from $10$\,$\mu$s to $800$\,$\mu$s. As an example, the data obtained for mode 3 is shown in Fig.\,\ref{fig:sbf_mode3}. Since for a 4-ion crystal the dynamics can be obtained numerically, we can obtain a temperature estimation by fitting the experimental data with the simulated curves of the sideband flops and searching for the optimal temperature values (shown in the right panel of Fig.\,\ref{fig:nbars}). Further description of the fitting method, together with the data of the other motional modes can be found in Appendix~\ref{app:rabi}.

For each pulse time below the cutoff, the global sideband ratio is calculated from $P_{r}$ and $P_{b}$ according to equation (\ref{eq:relation}) and individual estimates for $\bar{n}_i$ are obtained using equation (\ref{eq:maineq}), see left panel of Fig.\,\ref{fig:nbars}.

The cutoff time, as defined in Sec.\,\ref{subsec:theory_ion_cryst}, for all motional modes in this 4-ion crystal is around $400$\,$\mu$s ($gt\approx 0.22$). In order to avoid possible implicit biases in the single measurements of $\bar{n}_i$ at a given interrogation time, we use data from all available points up to the cutoff time to estimate the temperature. This set of $m=6$ individual $\bar{n}_i$ estimations, which are bias-corrected for $\N/2=200$ according to Eq.~(\ref{eq:crystal_bias}) and carry individual error bars $\sigma_i$, are averaged to obtain the final estimation according to the following weighted sum:
\begin{align}
    \hat{\bar{n}}=\arg\underset{\bar{n}}{\min}\sum_{i=1}^m\frac{(\bar{n}-\bar{n}_i)^2}{\sigma_i^2}.
    \label{eq:final_est}
\end{align}

The final estimation of $\hat{\bar{n}}$ for each motional mode is $\bar{n}=\{0.22\pm 0.05, 0.27\pm 0.02, 0.32 \pm 0.03, 0.35\pm 0.04\}$ and is shown in the right panel of Fig.\,\ref{fig:nbars}. The results are compared with the estimations obtained from the numerical fit of the temperature.

 Since a good agreement is found between the theoretical and experimental curves, the results from the fit are expected to give an accurate estimation for $\bar{n}$. For all the motional modes, the values of $\bar{n}$ obtained from the global sideband method agree with those extracted from the fit within a $1\sigma$ uncertainty. The global sideband ratio method reaches the same accuracy level of $\delta\bar{n}\sim 10-20\%$ as the numerical fit, but requires less data points to be taken. We want to explicitly emphasize that the used least-square fit relies on the exact calculation of the sideband dynamics, which scales exponentially with the number of ions and thus cannot be applied for larger ion crystals in general.
 
\section{Thermometry of a 2D crystal}
\label{sec:19ion_thermometry}

\subsection{Setup, state preparation and cooling}

\begin{figure}[t]
    \centering
    \includegraphics[width=.5\textwidth]{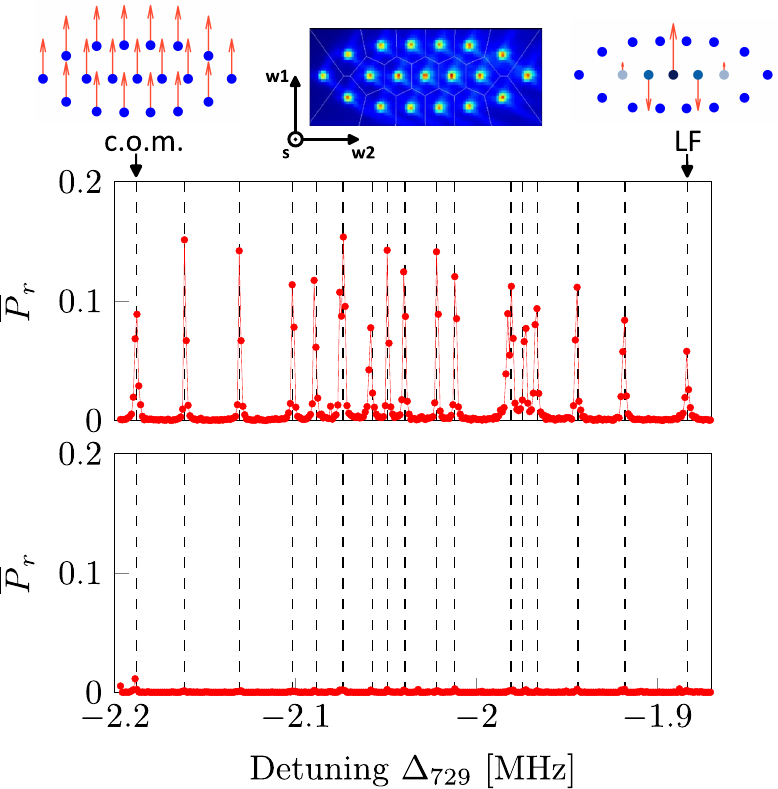}
    \caption{Red sideband spectrum of the out-of-plane modes of a 19-ion planar crystal after Doppler cooling (top) and an additional EIT-cooling laser pulse (bottom). The frequency is given as detuning from the $\ket{4\mathrm{S}_{1/2},~m=-1/2} \leftrightarrow \ket{3\mathrm{D}_{5/2},~m=-1/2}$ carrier transition. The mode frequencies obtained from simulations using pseudopotential theory are displayed as dashed lines. A false-color image of the ion crystal is shown at the top center. The mode structure of the COM mode (top left) as well as the lowest frequency mode (top right) is indicated by red arrows. Each arrow's length is proportional to the magnitude of the respective ion's Lamb-Dicke factor $\eta_i$, the direction indicating the sign of $\eta_i$. The temperature of both modes is probed in the experiments described further below.}
    \label{fig:19_spectrum}
\end{figure}

To demonstrate the new scheme also on a larger ion crystal, which already imposes challenges in numerically simulating the sideband dynamics, we perform thermometry measurements on the out-of-plane motional modes of a two-dimensional $^{40}$Ca$^+$ ion crystal. A planar 19-ion Coulomb crystal is stored in the anisotropic potential of a novel microfabricated monolithic linear Paul trap designed for trapping 2D ion crystals \cite{kiesenhofer2023}. The trap is operated at oscillation frequencies of $\omega_{\textrm{s}} = 2 \pi \times 2.189$ MHz, $\omega_{\textrm{w1}} = 2 \pi \times 645$ kHz and $\omega_{\textrm{w2}} = 2 \pi \times 340$ kHz where $\omega_{\textrm{s}}$ is the secular frequency in the strongly confining direction and $\omega_{\textrm{w1}}$ and $\omega_{\textrm{w2}}$ are the secular frequencies in the two weakly confining directions. The direction of the weakest confinement is aligned with the rf-zero line. Our geometry allows micromotion-free optical access in the plane spanned by the directions of $\omega_{\textrm{s}}$ and $\omega_{\textrm{w2}}$.

Ions are Doppler-cooled on the 4S$_{1/2}\leftrightarrow$ 4P$_{1/2}$ dipole transition at 397 nm. Electromagnetically-induced transparency (EIT) cooling \cite{morigi2000ground, roos2000experimental} is applied for 300 $\mu$s after Doppler cooling to simultaneously cool all $N$ out-of-plane secular modes of motion in an $N$-ion crystal close to the ground state. The direction of the magnetic field, defining the quantization axis, is oriented at an angle of 45$^\circ$ with respect to the crystal plane and allows for an optimal geometry for EIT cooling. A more detailed description of the beam geometry is given in Ref.~\cite{kiesenhofer2023}.

The laser used for EIT cooling is blue-detuned by 110\,MHz from the 4S$_{1/2}$ $\leftrightarrow$ 4P$_{1/2}$ transition. The chosen detuning enables efficient cooling over a frequency range large enough to accommodate all transverse modes of motion of a 19-ion crystal, spanning a range of $\sim 300$ kHz. The power of the $\sigma^{-}$-polarized beam is calibrated such that the induced Stark shift overlaps with the center of the frequency range to be cooled. Further details on the calibration procedure can be found in Ref.~\cite{Lechner2016}.

The motional modes are probed via sideband spectroscopy on the 4S$_{1/2} \leftrightarrow$ 3D$_{5/2}$ quadrupole transition with light from a frequency-stable laser ($\sim$~1~Hz linewidth) at 729 nm. A global beam along the out-of-plane direction excites the individual ions with a maximum variation in the single-ion carrier Rabi frequencies of about 6~\% across the 19-ion crystal.
For a single ion, we find $\bar{n}$ = 0.06 for the transverse mode of motion ($\omega_{\textrm{s}} = 2 \pi \times 2.188$ MHz) corresponding to the out-of-plane direction of a planar multi-ion crystal. A frequency scan of the red sideband spectrum of a planar 19-ion crystal, once after only Doppler cooling and once after an additional EIT cooling pulse of 300~$\mu$s, is shown in Fig.~\ref{fig:19_spectrum}.

\begin{figure*}[t]
    \centering
    \includegraphics[width=0.95\textwidth]{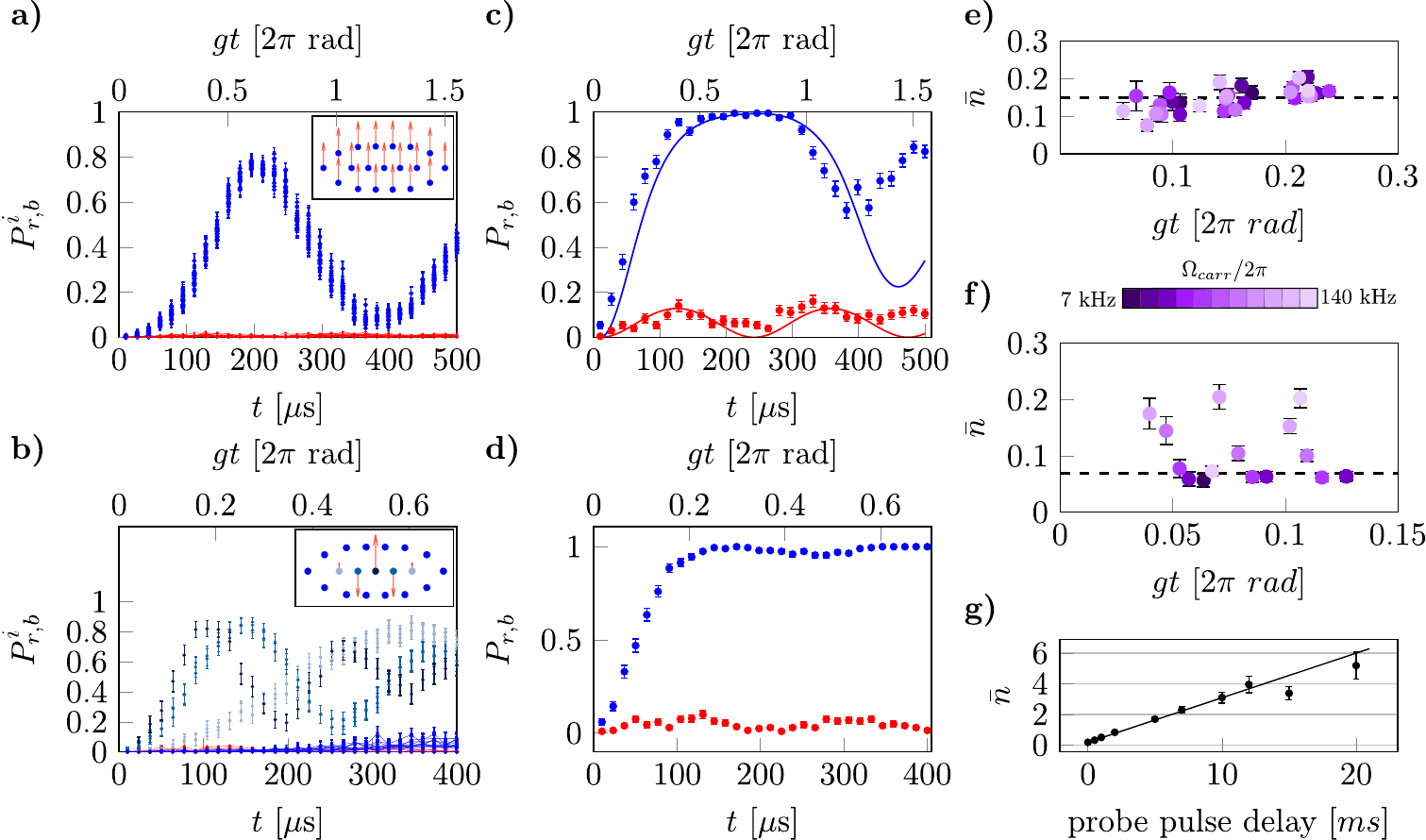}
    \caption{Sideband dynamics and thermometry of a planar 19-ion crystal. (a,b) Single-ion excitation probabilities $P_{r,b}^i$ for the COM mode (a) and the lowest-frequency mode (b). As in Fig.~\ref{fig:19_spectrum}, the insets indicate the mode vectors of the investigated modes. (c,d) Global excitation probabilities $P_{r,b}$ according to Eq.~\eqref{eq:pdef} for the COM mode (c) and the lowest-frequency mode (d) representing the quantities of interest in the sideband thermometry measurements. The blue and the red points correspond to the measured excitation on the blue and red sidebands, respectively. The solid lines in (c) are obtained from simulations in the symmetric Hilbert subspace using the least-square fitted value of $\bar{n}$.
    (e,f) Sideband thermometry of the COM mode (e) and the lowest-frequency mode (f) for varying carrier Rabi frequency (indicated by color) as a function of the interrogation time.
    The mean values are shown as dashed lines. For the lowest-frequency mode (f) the mean value is calculated after discarding all data points lying outside a range of $1 \sigma$ from the mean value obtained from all data points. The discarded data points (5 points with the highest values of $\bar{n}$) correspond to measurements with higher Rabi frequency and are biased due to crosstalk to neighbouring modes (see main text for discussion). The error bars of individual thermometry measurements are obtained from Eq.~(\ref{eq:crystal_var}).
    (g) Measurement of $\bar{n}$ for the COM mode as a function of the probe pulse delay. The solid line shows a weighted linear fit used to determine the heating rate.}
    \label{fig:thermometry_19_ions}
\end{figure*}

\subsection{Sideband thermometry of a planar 19-ion crystal}

Sideband thermometry based on Eq.~(\ref{eq:new_est}) is applied to the COM mode and the lowest-frequency mode in the out-of-plane direction of a two-dimensional 19-ion crystal.
We calculate the normal-mode frequencies and the Lamb--Dicke factors of the individual ions for all out-of-plane modes using simulations within the pseudopotential approximation. The knowledge of both is required for the employed temperature-estimation method. However, the pseudopotential approximation can sometimes fail to reproduce the observed normal mode frequency spectrum, in particular for planar crystals \cite{Landa2012,Kaufmann2012}. We measure the out-of-plane motional-mode spectrum and observe a good match with the simulated mode frequencies (see Fig.~\ref{fig:19_spectrum}), providing us with confidence that the pseudopotential approximation yields accurate results for the 19-ion crystal.

Sideband excitation dynamics of the COM and the lowest-frequency mode with a carrier Rabi frequency of $2 \pi \times 38.8$ kHz are shown in Fig.~\ref{fig:thermometry_19_ions}. Single-ion-resolved measurements reveal the structure of the investigated modes, shown as $P_{r,b}^i$ in Fig.~\ref{fig:thermometry_19_ions}(a,b), and yield the global excitation probabilities, $P_{b}$ and $P_{r}$, shown in Fig.~\ref{fig:thermometry_19_ions}(c,d).

Thermometry measurements are carried out after EIT cooling where we measure the temperature of the planar ion crystal as a function of the carrier Rabi frequency and the interrogation time. The excitation probabilities $P_{b}$ and $P_{r}$ are probed in 4000 individual experiments each. The mean phonon number is then calculated according to Eq.~(\ref{eq:new_est}). 

Using high laser intensities can lead to crosstalk to neighboring modes, which results in inaccuracies in the phonon-number estimates. For the lowest-frequency mode, we observe some crosstalk to the neighboring mode separated by less than 30~kHz. We can circumvent this problem simply by probing the modes at lower Rabi frequencies.
Figure~\ref{fig:thermometry_19_ions}(e,f) shows the estimated mean phonon numbers for different values of $gt$. Consistent results are obtained across the probed parameter space. The bias-corrected (Eq.~\ref{eq:crystal_bias}) weighted mean phonon number of the COM mode and lowest-frequency mode are determined to be $\bar{n} = 0.149(3)$ and $\bar{n} = 0.069(3)$, respectively.

To cross-check the measured mean phonon number of the COM mode, we simulate the sideband dynamics of the COM mode in the symmetric Hilbert subspace and, as for the 4-ion data, perform a weighted least-square fit of the measured data on the red sideband of the COM mode. More details on the numerical calculation for the symmetric COM mode as well as the fit estimator are given in Appendix~\ref{app:rabi}. The fit yields a temperature estimate of $\bar{n} = 0.147\pm0.02$, which is in good agreement with the value obtained using the new sideband thermometry technique. The theoretical curves for the red and the blue sideband are shown as solid lines in Fig.~\ref{fig:thermometry_19_ions}(c). Up to $\sim 350~\mu$s we find a good agreement between theory and experiment. For longer probe times, we observe deviations from the simulated curve, which we attribute to motional heating as well as instabilities in the trap oscillation frequencies due to power fluctuations. We thus use the first 20 data points ($<350~\mu$s) to fit the dynamics of the COM mode. A numerical simulation of the sideband dynamics of the LF mode, however, is computationally demanding. Therefore, a reference value for the lowest-frequency mode is not given.

In order to test the method with phononically higher excited states, a heating rate measurement of the COM mode is performed in which the probe pulse is delayed by a predefined wait time between 0 and 20~ms after ground-state cooling. The heating-rate curve in Fig.~\ref{fig:thermometry_19_ions}(g) shows the bias-corrected estimated values of $\bar{n}$. The data is fitted with a linear function by least-squares, weighted with the inverse variance obtained from Eq.~(\ref{eq:crystal_var}). The fit reveals a heating rate of 15.3(1.7) quanta/s per ion consistent with previous measurements with a single ion as well as an 8-ion crystal. In contrast to the COM mode, measurements on the lowest-frequency mode do not indicate significant heating within tens of milliseconds, as expected.
\section{Discussion \& Conclusions}
\label{sec:disc}

We have presented here a method for the thermometry of cold ICCs that generalizes the well-known sideband thermometry of single ions. The effects of many-body quantum dynamics that arise in this process can be taken into account with sufficient accuracy by exploiting conservation quantities of the sideband dynamics and by a suitable truncation of its time series expansion. It turns out that this limitation is not critical, since a sufficient signal can be extracted within a time span in which the truncation still yields reliable results. As we show, the tolerable interrogation time does not change with the number of ions. In principle, if required, a higher truncation order can also be achieved using the methods we have presented here, whose implementation in Python and Mathematica can be accessed at~\cite{REPO}. Applications of this methodology to a linear as well as a planar ion crystal give good results, also in comparison to other methods, in cases where such a comparison is possible. A reliable tool for temperature measurement in ultracold ion crystals is an important requirement in experimental quantum metrology and information science. We believe that the thermometry method presented here meets the current needs and can be of practical use for research with cold trapped ions. Moreover, the approach presented in this work could serve as a useful reference for the treatment of many-body effects in similar systems.

As an outlook, we would like to indicate a number of questions and possible further developments that go beyond the results presented here: a central premise of sideband thermometry is the presence of a canonical thermal state. This is a useful and mostly very good approximation, but it will not always be fulfilled for all cooling methods and especially not for short cooling durations, which will occur in quantum technology applications due to time limitations~\cite{Chen:2017,Rasmusson:2021}. Our general approach would also allow to consider more general, non-canonical parameterisations of the occupation probability and to estimate the corresponding parameters systematically. For this, one has to consider a many-parameter estimation problem and in a similar fashion derive the corresponding estimators from the measured observables. Another possible extension would be to consider correlated spin states in order to achieve a quantum metrological improvement of the accuracy of the thermometric measurement. Finally, a way could also be sought for thermometry based on bichromatic driving to exploit measurements of more than one ion and account for the many-body correlations that occur there.\\
\hfill\\
The data underlying the reported measurements are available via Zenodo~\cite{Zenodo}.

\section*{Acknowledgements}

The project leading to this application has received funding from the European Research Council (ERC) under the European Union’s Horizon 2020 research and innovation programme (grant agreement No 741541). Furthermore, we acknowledge funding from the Austrian Science Fund through the SFB BeyondC (F7110) and by the Deutsche Forschungsgemeinschaft (DFG, German Research Foundation) through Project-ID 274200144 – SFB 1227 (projects A06 and A07) and Project-ID 390837967 - EXC 2123. LD acknowledges support from the Alexander von Humboldt foundation.

\appendix
\section{Fisher information analysis for single ion thermometry}\label{app:FIzeros}

In this appendix, we analyse the single ion sideband temperature estimator in Eq.~\eqref{eq:sbratio} using the Fisher information formalism. The measured data (the excitation probabilities on the two sidebands) originate from a statistical model featuring an unknown parameter $\bar{n}$, which is to be determined using a certain estimator. The Fisher information $F(\bar{n})$ quantifies the amount of information about the unknown parameter that the chosen measurement scheme supplies. It connects to the variance of estimators $\Delta\bar{n}^2_{\textrm{err}}$ and the sample size $\N$ via the relation known as the Cramér-Rao inequality,
\begin{align}
    \Delta\bar{n}^2_{\textrm{err}}\geq \frac{1}{\N F(\bar{n})}\,.
    \label{eq:crineq}
\end{align}
The variance saturating this inequality is the Cramér-Rao bound, corresponding to the variance of the most efficient unbiased estimator for a given measurement scheme. It can be shown that in the single ion case the Cramér-Rao bound coincides with the quantum Cramér-Rao bound (QCRB), which takes into account all possible quantum measurements \cite{IVANOV2019101}.

\begin{figure}[tbp]
    \includegraphics{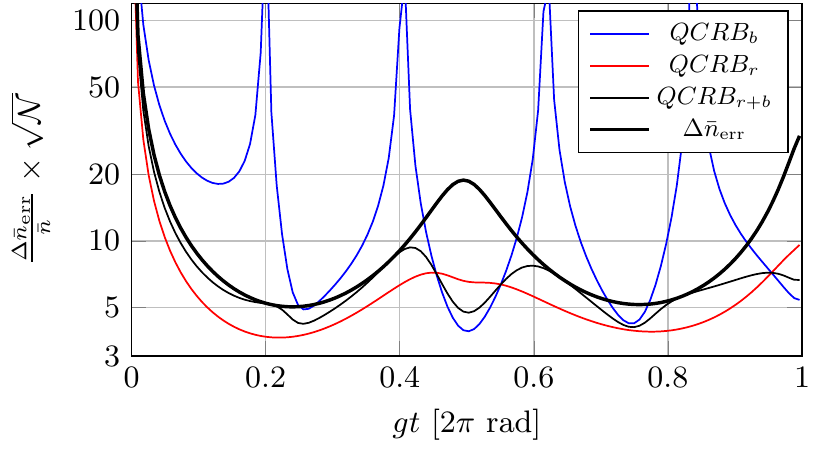}
    \caption[format=plain,justification=justified,singlelinecheck=false]{Statistical uncertainty of the temperature estimator rescaled to the sample size for different measurement schemes. Thick black curve represents the statistical uncertainty of the sideband estimator of Eq.~\eqref{eq:sbratio}. Thin curves are the quantum Cramér-Rao bounds for the three separate cases: when the measurements are taken on the red sideband (red curve), blue sideband (blue curve) and when both sidebands contribute equally to the data sample (gray). The figure is produced for $\bar{n}=0.1$.}
    \label{fig:crbapp}
\end{figure}

In Fig.~\ref{fig:crbapp}, we plot the statistical uncertainty $\Delta \bar{n}_{\textrm{err}}$ of the estimator given by Eq.~\eqref{eq:sbratio} as function of the sideband interrogation time together with the QCRB found from Eq.~\eqref{eq:crineq} for three separate measurement scenarios. The first case represents the situation with both sidebands equally contributing to the statistics, these are the conditions where formula~\eqref{eq:sbratio} is applied. The other two cases correspond to the situation when the data is gathered from only one of the sideband transitions, either red or blue.

From the plot one sees that the QCRB of the red sideband measurement almost everywhere lies below the other QCRB curves. This implies that the data gathered from the red sideband contain more information on the temperature than the data from the blue sideband or their equal combination. An estimator based solely on the red sideband only is thus potentially more efficient, yet for an absolute temperature determination it is not as practically convenient as the one of the sideband ratio estimator in Eq.~\eqref{eq:sbratio}, where the data from the blue sideband serves as an auto-calibration in presence of technical noise. One also sees that the sideband ratio estimator is a fairly efficient choice, since the corresponding statistical uncertainty lies mostly very close to the QCRB for the two-sideband scenario and coincides with it in multiple points.

The series of equidistant peaks in the blue sideband QRCB visible in Fig.~\ref{fig:crbapp} can be explained analytically. The red and blue sideband excitation probabilities for a single trapped ion are given by~\cite{RevModPhys.75.281}
\begin{subequations}
\begin{equation}
    P_r(\bar{n},t)=\frac{1}{2}\sum_{n=1}^{\infty}p_n(\bar{n})(1-\cos(gt\sqrt{n}))\,,
\end{equation}
\begin{equation}
    P_b(\bar{n},t)=\frac{1}{2}\sum_{n=0}^{\infty}p_n(\bar{n})(1-\cos(gt\sqrt{n+1}))\,,
\end{equation}
    \label{eq:flops_singleion}
\end{subequations}
respectively. Maxima of the Cramér-Rao bound correspond to minima of the Fisher information, which in the blue sideband case is given by
\begin{align}
\begin{split}
F_b(t,\bar{n})&=\frac{(\partial_{\bar{n}}P_b)^2}{P_b}+\frac{(\partial_{\bar{n}}(1-P_b))^2}{1-P_b}=\\
=&\qty[\frac{1}{P_b}+\frac{1}{1-P_b}]\qty(\sum_{n=0}^{\infty}\sin^2(t\sqrt{n+1})\partial_{\bar{n}}p_n(\bar{n}))^2\,.
        \label{eq:FI}
\end{split}
\end{align}
In the limit of $\bar{n}\ll 1$, the zeros of Fisher information are determined by the condition 
\begin{align}
   \sin^2(t)&=\sin^2(t\sqrt{2})\,,
   \label{eq:zerosFI}
\end{align}
for which we find the relevant solutions to be
\begin{align}
t&=2\pi\frac{k}{2(1+\sqrt{2})},\quad k\in \mathds{Z} \nonumber \\
&=2\pi\{0,0.207,0.414,0.621,0.828,\dots\}\,.
\end{align}
These solutions coincide well with the observed peaks of the blue sideband QCRB in Fig.~\ref{fig:crbapp}. This pattern of equidistant peaks emerges only in the low-temperature regime with $\bar{n}\ll 1$. 

\section{Sideband thermometry of ICC} \label{app:derivation}
The explicit form of the matrix elements $\bra{\mathbf{0},n}H_{r(b)}^{2k}\ket{\mathbf{0},n}$, needed for evaluating the power series of Eq.~\eqref{eq:series}, is found by direct evaluation. The spin and motional parts of expressions are split and evaluated separately, resulting in matrix elements of polynomials in $n$,
\begin{align*}
    \bra{\mathbf{0},n}H_{r}^0\ket{\mathbf{0},n}&=1\,,\\
    \bra{\mathbf{0},n}H_{b}^0\ket{\mathbf{0},n}&=1\,,\\
    \bra{\mathbf{0},n}H_{r}^2\ket{\mathbf{0},n}&=g^2\bra{\mathbf{0}}J_-J_+\ket{\mathbf{0}}\bra{n}a^{\dagger}a\ket{n}=g^2An\,,\\
    \bra{\mathbf{0},n}H_{b}^2\ket{\mathbf{0},n}&=g^2\bra{\mathbf{0}}J_-J_+\ket{\mathbf{0}}\bra{n}aa^{\dagger}\ket{n}=g^2A(n+1)\,,\\
    \bra{\mathbf{0},n}H_{r}^4\ket{\mathbf{0},n}&=g^4(B_2n(n-1)+B_1n^2)\,,\\
    \bra{\mathbf{0},n}H_{b}^4\ket{\mathbf{0},n}&=g^4(B_2(n+1)(n+2)+B_1(n+1)^2)\\
    ...
\end{align*}
Here the matrix elements are shown up to the fourth power of the Hamiltonians (see the supplementary Mathematica notebook for higher powers and the explicit expressions~\cite{REPO}). The coefficients $A$ and $B_i$ are obtained from the ground-state expectation values of certain strings of collective spin operators $J_{\pm}$. These coefficients could be efficiently computed analytically, which is discussed separately in Appendix~\ref{app:coeffs}. After plugging these expressions into the power series of Eq.~\eqref{eq:series} and averaging over the occuption number $n$, the sideband ratio (\ref{eq:relation}) takes the form of a polynomial in $\bar{n}$ and $t$,

\begin{equation}\label{eq:maineq_full}
\frac{P_r(t)}{P_b(t)-P_r(t)}\simeq \bar{n} + (gt)^2\mathcal{P}_2(\bar{n})-(gt)^4\mathcal{P}_3(\bar{n}) + (gt)^6\mathcal{P}_4(\bar{n}),
\end{equation}
as given in Eq.~\eqref{eq:mainresult}. The $k$-th order polynomials $\mathcal{P}_k(\bar{n})$ are 
\begin{equation*}
\mathcal{P}_2(\bar{n})=\frac{B_2}{6A}\bar{n}(1+\bar{n}),
\end{equation*}
\begin{widetext}
\begin{multline}
\mathcal{P}_3(\bar{n})=\frac{1}{360A^2}\bar{n}(1+\bar{n})(1+2\bar{n})\big[2(C_1+C_3+2C_4+3C_5)A-5B_2(2B_2+B_1)+15B_2A^2\big],\\
\mathcal{P}_4(\bar{n})=\frac{1}{30240A^3}\bar{n}(1 + \bar{n})\Big(-315A^4B_2(1 + 2\bar{n})^2 + 35B_2(B_1 + 2B_2)^2(1 + 2\bar{n})^2 + 42A^3(C_1 + C_3 + 2C_4 + 3C_5)(1 + 8\bar{n}(1 + \bar{n}))\\+
3A^2[12D_{1} + 2D_{10} + 3D_{11} + 2D_{12} + D_{13} + D_{14} + 9D_{3} + 6D_{4} + 3D_{5} + 2D_{6} + D_{7} + 6D_{8} + 4D_{9}+ \\+ 6(6D_{1} + 2D_{10} + 3D_{11} + 2D_{12} + D_{13} + D_{14} + 5D_{3} + 4D_{4} + 3D_{5} + 2D_{6} + D_{7} + 4D_{8} + 3D_{9})\bar{n}(1 + \bar{n}) -\\- 70(B_2 + 2B_2\bar{n})^2] - 14A\big\{B_1(C_1 + C_3 + 2C_4 + 3C_5)(1 + 2\bar{n})^2 + B_2\big(C_2 + 4(C_3 + 2C_4 + 3C_5) +\\+ 6(C_2 + 3C_3 + 5C_4 + 7C_5)\bar{n}(1 + \bar{n}) + 2C_1[2 + 9\bar{n}(1 + \bar{n})]\big)\big\}\Big).\label{eq:maineq_poly}
\end{multline} 
\end{widetext}

An appropriate root of Eq.~\eqref{eq:maineq_full} gives an estimator for the temperature of the motional mode. Although this equation generally has four roots, in practice it is easy to identify the one corresponding to the temperature estimation. The other roots are typically complex, negative or have values far outside of the $\bar{n}\lesssim 1$ region. If somehow the ambiguity is still present, dropping all time-dependent terms in the r.h.s. of ($\ref{eq:maineq_full}$) and measuring the l.h.s. at the smallest possible time provides the simplest rough estimation helping to choose the correct root. 

The asymptotic bias and the variance arising from the finite sampling of $\N/2$ for both $P_r$ and $P_b$ are calculated as:
\begin{align}
\begin{split}
    \delta\bar{n}&=\langle\hat{\bar{n}}\rangle-\bar{n}\\
    &=\frac{1}{\N}\Big[\frac{2P_bP_r(2-P_b-P_r)}{(P_b-P_r)^3}\frac{1}{\mathcal{R}_t'(\bar{n})}\\
    &\quad-\frac{P_bP_r(P_b+P_r-2P_bP_r)}{(P_b-P_r)^4}\frac{\mathcal{R}_t''(\bar{n})}{[\mathcal{R}_t'(\bar{n})]^3}\Big]\,,
    \label{eq:crystal_bias}
\end{split}
\end{align}
\begin{align}
    \Delta\hat{\bar{n}}_{\text{error}}^2=\frac{1}{\N}\frac{2P_bP_r(P_b+P_r-2P_bP_r)}{(P_b-P_r)^4}\frac{1}{[\mathcal{R}_t'(\bar{n})]^2}\,,
    \label{eq:crystal_var}
\end{align}
where the derivative of $\mathcal{R}_t(\bar{n})$ is understood as the derivative with respect to $\bar{n}$.

\section{Mode-dependent coefficients}\label{app:coeffs}
The coefficients $A$, $B_i$, $C_i$ and $D_i$ given in Eqs.~\eqref{eq:maineq_poly} depend only on the structure of the interrogated motional mode and are defined as follows:
\begin{align*}
A&=\bra{\mathbf{0}}J_-J_+\ket{\mathbf{0}}=\sum_{i=1}^N\eta_i^2=1,\\
B_1&=\bra{\mathbf{0}}(J_-J_+)^2\ket{\mathbf{0}}=\qty(\sum_{i=1}^N\eta_i^2)^2=1,\\
B_2&=\bra{\mathbf{0}}J_-^2J_+^2\ket{\mathbf{0}}=2\qty(\qty(\sum_{i=1}^N\eta_i^2)^2-\sum_{i=1}^N\eta_i^4)\,,
\end{align*} 
\begin{align*}
C_1 &=\bra{\mathbf{0}}J_-J_-J_+J_+J_-J_+\ket{\mathbf{0}}\,,\\
C_2 &=\bra{\mathbf{0}}J_-J_+J_-J_+J_-J_+\ket{\mathbf{0}}\,, \\
C_3 &=\bra{\mathbf{0}}J_-J_+J_-J_-J_+J_+\ket{\mathbf{0}}\,, \\ 
C_4 &=\bra{\mathbf{0}}J_-J_-J_+J_-J_+J_+\ket{\mathbf{0}}\,,\\
C_5 &=\bra{\mathbf{0}}J_-J_-J_-J_+J_+J_+\ket{\mathbf{0}}\,,
\end{align*} 
\begin{align*}
D_1 &=\bra{\mathbf{0}}J_-J_-J_-J_-J_+J_+J_+J_+\ket{\mathbf{0}}\,,\\
D_2 &=\bra{\mathbf{0}}J_-J_+J_-J_+J_-J_+J_-J_+\ket{\mathbf{0}}\,,\\ 
D_3 &=\bra{\mathbf{0}}J_-J_-J_-J_+J_-J_+J_+J_+\ket{\mathbf{0}}\,,\\ 
D_4 &=\bra{\mathbf{0}}J_-J_-J_+J_-J_-J_+J_+J_+\ket{\mathbf{0}}\,,\\
D_5 &=\bra{\mathbf{0}}J_-J_+J_-J_-J_-J_+J_+J_+\ket{\mathbf{0}}\,,\\ 
D_6 &=\bra{\mathbf{0}}J_-J_-J_+J_+J_-J_-J_+J_+\ket{\mathbf{0}}\,,\\ 
D_7 &=\bra{\mathbf{0}}J_-J_+J_-J_+J_-J_-J_+J_+\ket{\mathbf{0}}\,,\\
D_8 &=\bra{\mathbf{0}}J_-J_-J_-J_+J_+J_-J_+J_+\ket{\mathbf{0}}\,,\\
D_9 &=\bra{\mathbf{0}}J_-J_-J_+J_-J_+J_-J_+J_+\ket{\mathbf{0}}\,,\\ 
D_{10} &=\bra{\mathbf{0}}J_-J_+J_-J_-J_+J_-J_+J_+\ket{\mathbf{0}}\,,\\
D_{11} &=\bra{\mathbf{0}}J_-J_-J_-J_+J_+J_+J_-J_+\ket{\mathbf{0}}\,,\\
D_{12} &=\bra{\mathbf{0}}J_-J_-J_+J_-J_+J_+J_-J_+\ket{\mathbf{0}}\,,\\ 
D_{13} &=\bra{\mathbf{0}}J_-J_+J_-J_-J_+J_+J_-J_+\ket{\mathbf{0}}\,,\\
D_{14} &=\bra{\mathbf{0}}J_-J_-J_+J_+J_-J_+J_-J_+\ket{\mathbf{0}}\,.
\end{align*}
Essentially, the coefficients are all the non-vanishing expectation values of strings of operators $J_{\pm}$ of a fixed length. Using the definition $J_{\pm}=\sum_{i=1}^N\eta_i\sigma_i^{\pm}$ each of the coefficients is decomposed into a sum of many strings of particular single-atom Pauli operators. Each string is weighted with the corresponding prefactor, consisting of the mode vector components and sandwiched with the spin ground state. With some combinatorics one can classify and pick out the small number of non-zero terms to ease the calculation. There may be several ways to do so, one of them would be to distinguish the terms based on the number of unique atomic indices appearing in an individual term. For each mode-dependent coefficient we count the number of terms of each class ($C_i^j$, $D_i^j$) and then multiply it with the corresponding expectation values. This brings us to the general expression for $C$- and $D$-coefficients given by
\begin{align}
    C_i=&C_i^1\sum_{i=1}^N\eta_i^6+C_i^2\sum_{i=1}^N\sum_{j\neq i}^N\eta_i^4\eta_j^2+C_i^3\sum_{i=1}^N\sum_{j\neq i}^N\sum_{k\neq i,j}^N\eta_i^2\eta_j^2\eta_k^2, \label{eq:ccoeffs}
\end{align}
\begin{align}
    D_i=&D_i^1\sum_{i=1}^N\eta_i^8+D_i^2\sum_{i=1}^N\sum_{j\neq i}^N\eta_i^6\eta_j^2+D_i^3\sum_{i=1}^N\sum_{j\neq i}^N\eta_i^4\eta_j^4+\nonumber \\&+D_i^4\sum_{i=1}^N\sum_{j \neq i}^N\sum_{k\neq i,j}^N\eta_i^4\eta_j^2\eta_k^2+\nonumber \\&+D_i^5\sum_{i=1}^N\sum_{j\neq i}^N\sum_{j\neq i}^N\sum_{k\neq i,j}^N\sum_{p\neq i,j,k}^N\eta_i^2\eta_j^2\eta_k^2\eta_p^2\,.\label{eq:dcoeffs}
\end{align}
Listed in the tables \ref{tab:t1} and \ref{tab:t2} are the prefactors completing the expressions (\ref{eq:ccoeffs}, \ref{eq:dcoeffs}), which are needed to evaluate each of the $C$- and $D$-coefficients.

\begin{table}[htbp]
  \centering
  \caption{$C_i^j$-coefficients}
    \begin{tabular}{|c||c|c|c|}
    \hline
         $i$ & \multicolumn{1}{l|}{$C_i^1$} & \multicolumn{1}{l|}{$C_i^2$} & \multicolumn{1}{l|}{$C_i^3$} \bigstrut\\
    \hline\hline
    $1$ & 0     & 4     & 2 \bigstrut\\
    \hline
    $2$ & 1     & 3     & 1 \bigstrut\\
    \hline
    $3$ & 0     & 4     & 2 \bigstrut\\
    \hline
    $4$ & 0     & 4     & 4 \bigstrut\\
    \hline
    $5$ & 0     & 0     & 6 \bigstrut\\
    \hline
    \end{tabular}%
  \label{tab:t1}%
\end{table}%

\begin{table}[htbp]
  \centering
  \caption{$D_i^j$-coefficients}
    \begin{tabular}{|c||c|c|c|c|c|}
    \hline
      $i$ & \multicolumn{1}{l|}{$D_i^1$} & \multicolumn{1}{l|}{$D_i^2$} & \multicolumn{1}{l|}{$D_i^3$} & \multicolumn{1}{l|}{$D_i^4$} & \multicolumn{1}{l|}{$D_i^5$} \bigstrut\\
    \hline\hline
    $1$ & 0     & 0     & 0     & 0     & 24 \bigstrut\\
    \hline
    $2$ & 1     & 4     & 3     & 6     & 1 \bigstrut\\
    \hline
    $3$ & 0     & 0     & 0     & 18    & 18 \bigstrut\\
    \hline
    $4$ & 0     & 0     & 0     & 24    & 12 \bigstrut\\
    \hline
    $5$ & 0     & 0     & 0     & 18    & 6 \bigstrut\\
    \hline
    $6$ & 0     & 0     & 8     & 16    & 4 \bigstrut\\
    \hline
    $7$ & 0     & 4     & 4     & 10    & 2 \bigstrut\\
    \hline
    $8$ & 0      & 0      & 0      & 24      & 12 \bigstrut\\
    \hline
    $9$ & 0      & 4      & 4      & 24      & 8 \bigstrut\\
    \hline
    ${10}$ & 0      & 4      & 4      & 16      & 4 \bigstrut\\
    \hline
    ${11}$ & 0      & 0      & 0      & 18      & 6 \bigstrut\\
    \hline
    ${12}$ & 0      & 4      & 4      & 16      & 4 \bigstrut\\
    \hline
    ${13}$ & 0      & 4      & 4      & 10      & 2 \bigstrut\\
    \hline
    ${14}$ & 0      & 4      & 4      & 10      & 2 \bigstrut\\
    \hline
    \end{tabular}%
  \label{tab:t2}%
\end{table}%

In order to use the new temperature estimation method, for a given motional mode one needs to compute in total 22 mode-dependent coefficients. Using formulas (\ref{eq:ccoeffs}, \ref{eq:dcoeffs}), the calculation boils down to programming a cascade of several FOR-loops, which results in a computationally-friendly polynomial scaling with respect to $N$. Hence, one may calculate the coefficients entering the estimator easily and fast to apply the new thermometry method for arbitrarily large ion crystals. We provide a program for calculating these coefficients for a given motional mode in the code supplement~\cite{REPO}.

\section{Numerical fit estimator} \label{app:rabi}
\begin{figure}[tbp]
    \centering
    \includegraphics{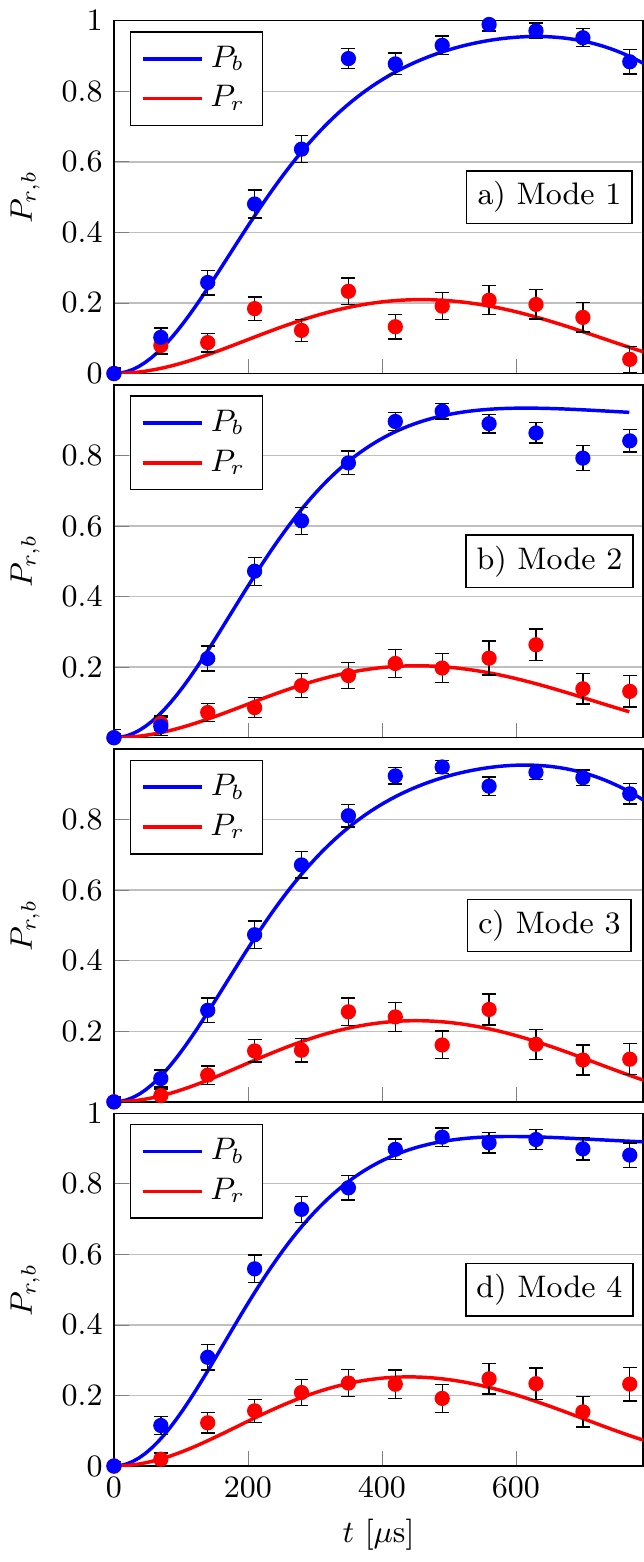}
    \caption{Red (lower points) and blue (upper points) global sideband flops of all 1-4 motional modes shown in (a)-(d), respectively, for the 4-ion crystal. The solid lines show the exact numerical solution, where $\bar{n}$ was fitted as free parameter with the weighted least-square method.}
    \label{fig:sbf_all_modes}
\end{figure}

A common way for estimating cold ion temperatures is fitting the experimental data with theory curves while using the temperature as a free parameter. Naturally, this technique requires the exact numerical solutions of the Schrödinger equation or its reasonable approximation to be available. This is, however, problematic for large ion crystals due to the exponential scaling of the Hilbert space, and causes the main bottleneck for the numerical fit. For smaller crystals, one can employ this method for temperature estimation using a weighted least square optimization of the model curves with respect to unknown parameter $\bar{n}$. The choice of model functions may vary, though in our case the best performance was achieved when using the red sideband collective excitation probability $P_r(t,\bar{n})$. This choice is further backed by the previous observation from the single-ion case, that the Fisher information on parameter $\bar{n}$ contained in the red sideband is typically significantly higher than the one contained in the blue sideband or in the combinations of both. This observation also holds in the multi-ion case for relatively short interrogation times. Using the red sideband curves as model functions makes the numerical fit estimator in our case have the following form
\begin{align}
    \hat{\bar{n}}=\arg\underset{\bar{n}}{\min}S(\bar{n})=\arg\underset{\bar{n}}{\min}\sum_{i=1}^m\frac{(P_r(t_i,\bar{n})-x_i)^2}{\sigma_i^2}\,,
    \label{eq:leastsq}
\end{align}
where the sum is taken over $m$ experimental points $x_i$, each carrying a normally-distributed error $\sigma_i$. The variance of this non-linear weighted least-square estimator is given by \cite{CLINE1970291}
\begin{align}
    \Delta\bar{n}^2_{\textrm{err}}&=\qty[\sum_{i=1}^m\qty(\frac{A_i}{\sigma_i})^2]^{-1}\qty(S_L-S(\hat{\bar{n}}))\,,\\
    S_L&=S(\hat{\bar{n}})\qty(1+\frac{1}{m-1}F(1,m-1,1-\beta))\,,
\end{align}
with $A_i=\partial P_r(t_i,\bar{n})/\partial \bar{n}$ and $F(d_1,d_2,1-\beta)$ being the quantile function of statistical F-distribution with parameters $d_1, d_2$, taken at a point $1-\beta$. The value $\beta=0.317$ defines a confidence interval of one standard deviation and is used for the calculations. For the experiment with a 4-ion crystal discussed in Section~\ref{sec:4ion_thermometry}, this method was used to fit the obtained sideband excitation data. The results of the numerical fit together with the experimental data for all motional modes are shown in Fig.~\ref{fig:sbf_all_modes}.

The symmetric center-of-mass mode deserves a special mention. Since all the individual coupling strengths are equal for this mode ($\eta_i=1/\sqrt{N}\: \forall i$), the spin dynamics evolves within the symmetric Hilbert subspace and could effectively be described using symmetric spin Dicke basis of states
\begin{align}
\ket{\mathbf{M}}&=\frac{1}{M!}{\binom{N}{M}}^{-1/2}S_+^M\ket{\mathbf{0}}, & S_+&=\sum_{i=1}^N\sigma_i^+\,.
\end{align}
This lifts the exponential scaling of the Hilbert space with respect to the number of ions $N$ and thus makes the numerical fit applicable for the center of mass mode of larger ion crystals.

\end{document}